\documentclass{aa}
\usepackage{graphicx,amsmath}
\usepackage{txfonts}

\usepackage{natbib} 
\bibpunct{(}{)}{;}{a}{}{,}
\begin{document}
\topmargin-3cm
 \title{The magnetic flux of the quiet Sun internetwork as observed with the Tenerife Infrared Polarimeter}

   \author{C. Beck\inst{1} \and R. Rezaei\inst{2} }
        
   \titlerunning{Magnetic flux in the internetwork}
  \authorrunning{C. Beck et al.}  
\offprints{C. Beck}

   \institute{Instituto de Astrof\'{\i}sica de Canarias\\
      \and Kiepenheuer-Institut f\"ur Sonnenphysik\\
        }
 
\date{Received xxx; accepted xxx}

\abstract{Recent observations with the spectropolarimeter on-board the HINODE satellite have found abundant horizontal magnetic fields in the internetwork quiet Sun.}{We compare the results on the horizontal fields obtained at 630\thinspace nm with ground-based observations at 1.56 $\mu$m where the sensitivity to magnetic fields is larger than in the visible.}{We obtained 30 sec-integrated spectropolarimetric data of quiet Sun on disc centre during a period of extremely stable and good seeing. The data have a rms noise in polarization of around $2\cdot 10^{-4}$ of the continuum intensity. The low noise level allowed for an inversion of the spectra with the SIR code. We compare the inversion results with proxies for the determination of magnetic flux.}{We confirm the presence of the horizontal fields in the quiet Sun internetwork as reported from the satellite data, with voids of some granules extent of nearly zero linear polarization signal. Voids in the circular polarization signal are only of granular scale. More than 60 \% of the surface show polarization signals above four times the rms noise level. We find that the total magnetic flux contained in the more inclined to horizontal fields ($\gamma > 45^\circ$) is smaller by a factor of around 2 than that of the less inclined fields. The proxies for flux determination are seen to suffer from a strong influence of the thermodynamic state of the atmosphere, and hence, seem to be unreliable.}{During spells of good seeing conditions, adaptive optics can render ground-based slit-spectrograph observations at a 70-cm telescope equivalent to the seeing-free space-based data of half-meter class telescopes. We suggest that the difference of the ratio of horizontal to transversal flux between the ground-based infrared data and the satellite-based visible data is due to the different formation heights of the respective spectral lines. We caution that the true amount of magnetic flux cannot be derived directly from the spectra. For purely horizontal flux, one would need its vertical extension that has to estimated by an explicit modeling with the observed spectra as boundary conditions, or has to be taken from MHD simulations. Time-series of the evolution of the magnetic flux and chromospheric diagnostics are needed to address its possible contribution to chromospheric heating.}
\maketitle
\section{Introduction}
Recent studies using  observations of magnetic fields in quiet Sun regions obtained with the  spectropolarimeter on-board the HINODE satellite \citep{kosugi+etal2007} have found abundant horizontal magnetic fields \citep[][; OR07 and LI08 in the following]{orozco+etal2007,lites+etal2008}, whose existence was predicted before from the mismatch of observed average magnetic flux between Zeeman and Hanle measurements by \citet{trujillobueno+etal2004}. Studies of photospheric magnetic fields using Zeeman-sensitive spectral lines in the visible (VIS) wavelength range, however, suffer from a severe drawback inherent to the wavelength: the thermal broadening ($\propto\lambda$) is dominating over the Zeeman splitting due to magnetic field ($\propto\lambda^2$). For spectral lines in the near-infrared (IR), the relation is more favorable. The retrieval of magnetic field properties from spectra in the VIS thus is then less reliable, because the lines are in the weak-field limit that imposes severe limits \citep{marian+etal2006,marian+etal2008a}. The separation of the oppositely polarized spectral components is not proportional to the field strength, and the information on field strength and magnetic flux both influence the amplitude of the polarization signal in similar ways. Observations of solar magnetic fields commonly do not resolve the magnetic structures, thus, the polarization signal amplitude, which is always measured relative to intensity, gets further complicated by the question of spatial resolution, stray light and the thermodynamics of the solar atmosphere \citep{marian+etal2006}. An accurate determination of magnetic field properties from the weak polarization signals prevalent in the quiet Sun in VIS lines thus gets strongly dependent on the quality of the observations, on the signal-to-noise ratio \citep{bellotrubio+collados2003,rezaei+etal2007} and the spatial resolution \citep{khomenko+etal2005a}. IR spectra at 1.56 $\mu$m allow to disentangle most of the ambiguities because the field strength can be read off from the splitting of the polarization, or in some cases, even the intensity components of the spectral line for magnetic fields above around 400 G \citep{beck2006,beck+etal2007}. It is thus not surprising that studies before HINODE using solely VIS lines  \citep{almeida+lites2000,cerdena+etal2003,navarro+almeida2003,lites+socke2004,socke+lites2004} or IR lines \citep{lin1995,khomenko+etal2003,marian+etal2006a,cerdena+etal2006} did not agree especially on the question of the intrinsic field strength: the VIS observations indicated a significant fraction of strong kG fields, whereas the IR was dominated by weaker hG fields. 
\begin{figure*}
$ $\\$ $\\
\centerline{\resizebox{17cm}{!}{\includegraphics{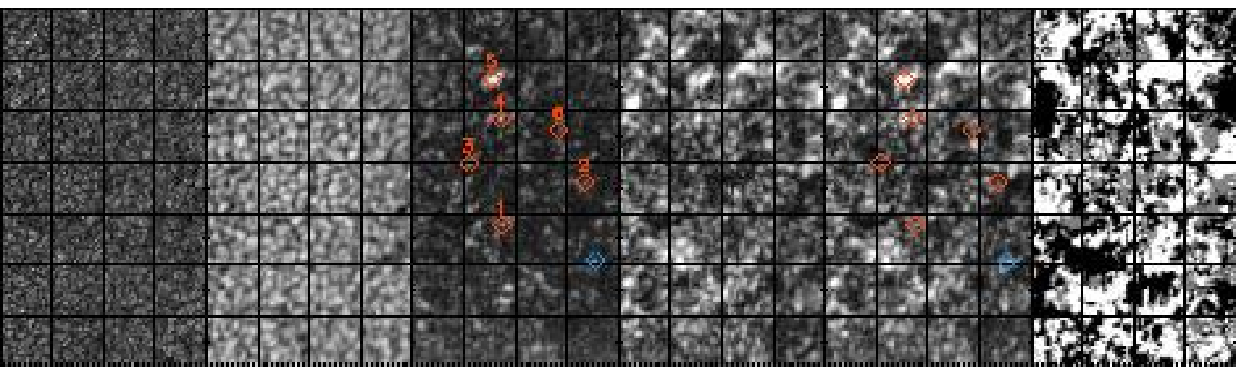}}}$ $\\$ $\\$ $\\
\centerline{\resizebox{17cm}{!}{\includegraphics{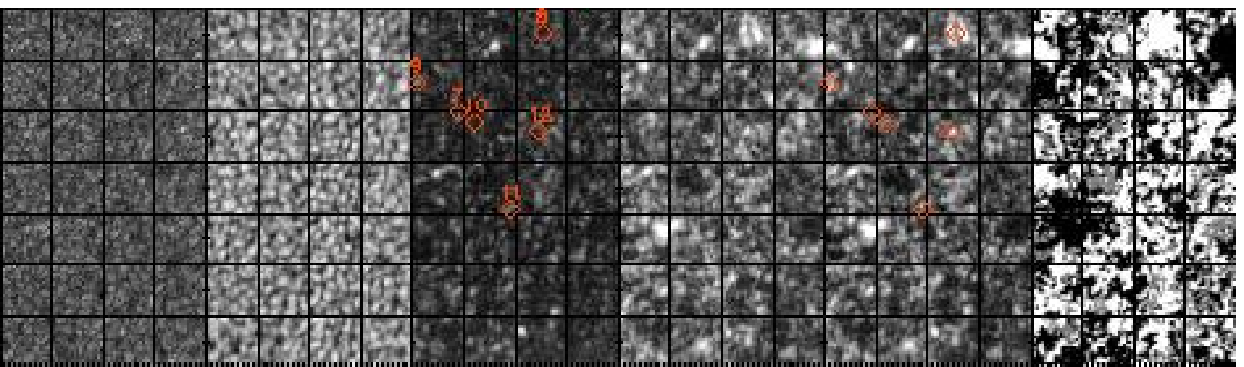}}}$ $\\
\caption{Overview maps of the observations on 21.5.08. {\em Top}: TIP Op.~001,
  taken at around UT 8:30. {\em Bottom}: TIP Op.~005, from around UT
  9:30. {\em Left to right}: G-band image, $I_c$,  $L_{max}$,  $V_{max}$, $p$, polarity. Scale is in arcsec, grid lines have a 10 arcsec spacing. { The {\em blue diamonds} denote the location of the profile shown in Fig.~\ref{fig10}, the {\em red diamonds} the locations of those shown in Appendix \ref{appa}.}\label{fig1}}
\end{figure*}

The contradictory findings from VIS or IR spectral lines could only recently be reconciled to some degree. \citet{rezaei+etal2007} retrieved for the first time a predominance of weak fields in the internetwork (IN) from observations at 630 nm in a study on the relation between photospheric fields and chromospheric emission. They used ground-based observations that were obtained using the Kiepenheuer-Institute adaptive optics system \citep{vdluehe+etal2003} and the POLIS spectropolarimeter \citep{beck+etal2005b} at the German Vacuum Tower Telescope (VTT) in Iza{\~n}a. \citet{marian+etal2008a} showed that for simultaneous observations in VIS and IR lines taken under identical seeing conditions, the VIS observations are compatible with the weak fields retrieved from the corresponding IR spectra. The VIS spectra alone were, however, found again to be biased towards kG fields due to the limitations imposed by the weak-field limit. 

The new findings on the magnetic fields in the internetwork quiet Sun from the HINODE in VIS lines thus should again be best compared and confirmed by corresponding observations in the IR. During a recent observation campaign, we were able to obtain IR spectra at 1.56$\mu$m during a period of extremely good and stable seeing, yielding a nearly uniform spatial resolution in time and space that was near the diffraction limit of the VTT at these wavelengths (0\farcs6). We describe the observations in Sect.~\ref{obs} and  explain additional analysis steps and the inversion of the spectra with the SIR code \citep{cobo+toroiniesta1992} in Sect.~\ref{ana}. The results are presented in Sect.~\ref{res} and summarized and discussed in Sect.~\ref{disc}. Our conclusions are given in Sect.~\ref{concl}.
\section{Observations\label{obs}}
On 21/05/2008, we observed a region of quiet Sun on disc centre with the
Tenerife Infrared Polarimeter \citep[TIP;][]{martinez+etal1999,collados+etal2007} at the German
Vacuum Tower Telescope (VTT) in Iza{\~n}a. We used the 1.56
$\mu$m range that includes two Zeeman sensitive { \ion{Fe}{I}} solar
spectral lines at 1564.8 nm { ($g_{eff}=3$)} and 1565.2 nm { ($g_{eff}=1.5$)}. The slit width of 100 $\mu$m corresponded to
0\farcs45. The slit was stepped in 0\farcs5 steps across the solar image in
the focal plane. The spatial sampling along the slit was 0\farcs175; the
spectra were binned later to a 0\farcs35 sampling that is sufficient to sample the
diffraction limit of the VTT (0\farcs6 at 1.5 $\mu$m). Due to the extremely
good and especially stable seeing conditions, we used an integration time of
30 secs per scan step. Two maps of
40$^{\prime\prime}$x70$^{\prime\prime}$ were taken  from UT 7:58 until 8:43 (Operation 001), and from  UT 9:16 until 10:01 (Operation 005).  The TIP
data were reduced with the respective routines
\citep[e.g.,][]{schlichenmaier+collados2002} including the correction for the telescope polarization \citep{beck+etal2005a}. The noise level of the
polarization signal in continuum windows is around 2$\cdot 10^{-4}$ of the
continuum intensity. 

We used a narrow-band G-band imaging channel for speckle-reconstructed
context images and the TElecentric SOlar Spectrometer
\citep[TESOS;][]{kentischer+etal1998} for G-band spectroscopy simultaneously
with TIP. We aligned the speckle-reconstructed G-band images taken during the
TIP maps with the same methods as described in \citet{beck+etal2007}; the
TESOS spectra were not considered for the present study.

Figure \ref{fig1} shows overview maps of the observations: the aligned speckle-reconstructed G-band data, the continuum intensity $I_c$ at around 1565 nm, the maximum linear polarization signal $L_{max}$ of the 1564.8 nm line, the maximum circular polarization signal $V_{max}$, the polarization degree $p$, and finally, the polarity of the magnetic fields that indicates if the field lines were
(anti)parallel to the line of sight (LOS). { The polarity was defined as
  $\pm$1 ({\em black} and {\em white} in Fig.~\ref{fig1}) from the order of
  minimum and maximum Stokes $V$ signal around the 1564.8 nm line for all
  locations with a polarization degree above the threshold discussed
  below. Locations without significant polarization signal appear {\em grey}
  in the polarity map.} $L_{max}$, $V_{max}$ and $p$
 were defined as the largest polarization signal in a small wavelength range
around the line core of 1564.8 nm in $L=\sqrt{Q^2+U^2}$, Stokes $V$, and
$P=\sqrt{Q^2+U^2+V^2}$, respectively. Assuming that the speckle-reconstructed
G-band data ({\em leftmost image}) is a close representation of the true solar
surface structure, the TIP data show the same structure only worsened by the
limitation of the spatial sampling, despite the 30 sec integration. The cadence
of the G-band data was interrupted in both observations near the end of the
scanning, { and the G-band data has no exact timing information. It was
  thus impossible to obtain a good alignment to $I_c$ for the last
  approximately 10$^{\prime\prime}$ of the scans.} $L_{max}$,
$V_{max}$ and $p$ are all displayed with an identical upper threshold of 1
\% of $I_c$ of polarization amplitude. $p$ is dominated by the contribution from the
circular polarization, i.e., nearly vertical magnetic fields, but there are
several locations with strong linear polarization signal as well. Note that
only a small fraction of the magnetic fields leads to bright points in the
G-band (e.g., x/y= 8-15/25-20$^{\prime\prime}$ in the {\em upper left} image), like also found by \citet{dewijn+etal2008}.
The second map ({\em lower row}), taken around an hour later, shows again the strongest and stable magnetic flux concentrations of the previous observation displaced by around 5$^{\prime\prime}$: the black patch in the polarity map at x/y = 38/55$^{\prime\prime}$ or the small white patch at 18/58$^{\prime\prime}$ can be re-identified in the 2nd map as well at 38/65$^{\prime\prime}$ and
18/63$^{\prime\prime}$, respectively.
\begin{figure}
$ $\\$ $\\
\centerline{\resizebox{7.8cm}{!}{\includegraphics{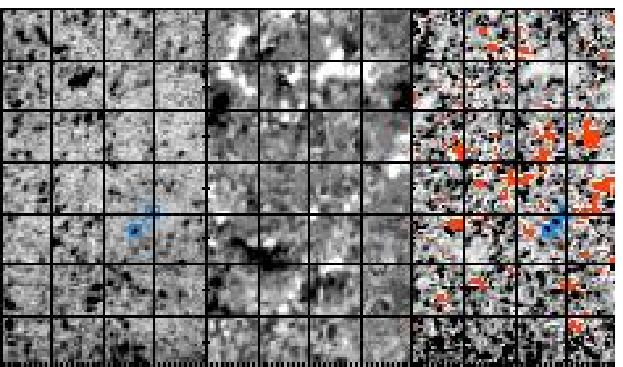}}}$ $\\$ $\\$ $\\
\centerline{\resizebox{7.8cm}{!}{\includegraphics{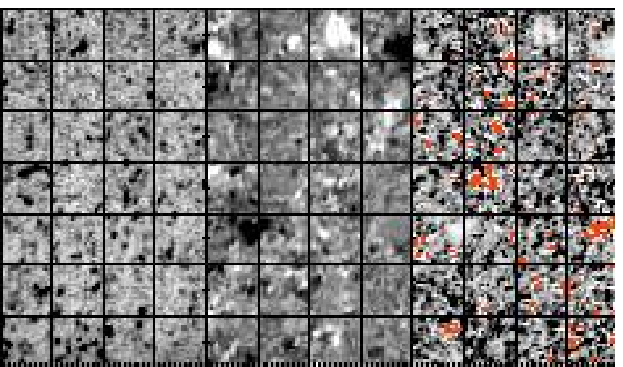}}}$ $\\
\caption{{\em Left to right}: $L_{tot}$, $V_{tot}$ times polarity,
  ratio $L/V$. The {\em red patches} in $L/V$ denote locations without any clear $V$ signal to divide with. First and last map are displayed inverted. The {\em blue rectangles} in the {\em upper left image} denote two patches of large $L_{tot}$, between which in $L/V$ ({\em upper right image}) another one shows up. {\em Top/bottom}: TIP Op.~001 and Op.~005. Scale is in arcsec, grid lines have a 10 arcsec spacing.\label{fig3}}
\end{figure}
\section{Data analysis\label{ana}}
For a comparison with the study of LI08, we also calculated
the total linear and circular polarization, $L_{tot}$ and $V_{tot}$, by an
integration of the absolute polarization signal $L(\lambda)$ and
$|V(\lambda)|$ over wavelength. These quantities are related to the
longitudinal ($B_L$) and transversal ($B_T$) component of the magnetic
field. The (anti)parallel direction of the longitudinal fields is introduced
into $V_{tot}$ again by a multiplication with the polarity map. Figure
\ref{fig3} displays maps of $L_{tot}$ and $V_{tot}$. We added additionally
maps of the ratio of linear and circular polarization, $L/V$. The ratio was
calculated by an average over a few spectral pixels ($\sim$10 pm) near the location of
the largest $V$ signal of 1564.8 nm \citep[see][]{beck2006}. The spatial  pattern in
these maps is strikingly similar to those found by LI08:
the polarization signal shows ``voids'' of some granules extent (about 10$^{\prime\prime}$ diameter in $L_{tot}$, 3-5$^{\prime\prime}$in $V_{tot}$) that are free from any clear signal; the linear polarization signal appears in the shape of
small ``blobs'' around the edges of the voids. The ``blobs'' of enhanced $L_{tot}$  are slightly smaller than individual granules \citep[cf.][]{ishikawa+etal2008}. In the ratio $L/V$ ({\em
  right column}), several more locations with stronger linear than circular
polarization show up in the form of similar ``blobs'' as in $L_{tot}$. For example, at x/y=26/26$^{\prime\prime}$ and 30/30$^{\prime\prime}$ in the {\em upper left image} two ``blobs'' appear in $L_{tot}$ ({\em blue rectangles}), whereas in $L/V$ ({\em upper right}) a third one is additionally seen right between them at x/y=28/28$^{\prime\prime}$.
\begin{figure}
\centerline{\hspace*{.3cm}\resizebox{4.cm}{!}{\includegraphics{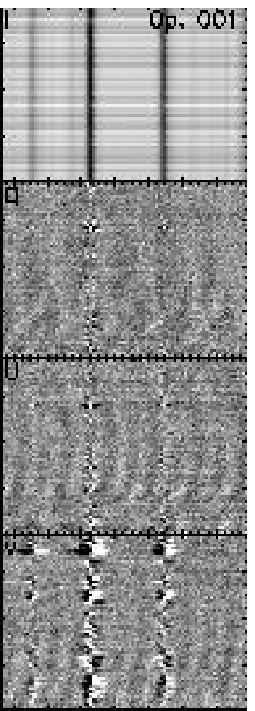}}\resizebox{4.cm}{!}{\includegraphics{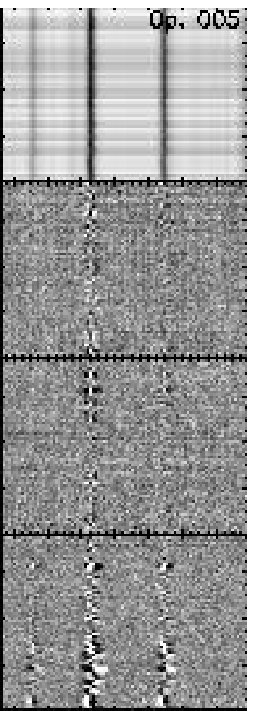}}}$ $\\
\caption{{ Example of {\em top to bottom}} $IQUV$ spectra for one scan step from Op.~001 {{\em at left}} and  Op.~005  {{\em at right}}. The display threshold for $QUV$ was set to $\pm$0.001 $I_c$ .\label{fig2}}
\end{figure}

Figure \ref{fig2} shows one example of slit-spectra from each observation. There is a residual (vertical) fringe pattern in $QUV$ of
Op.~001 that comes from imperfect flatfielding{ , with a maximum amplitude of around 0.0005 of $I_c$ in Stokes $U$; the amplitude is smaller in $Q$ and $V$. The amplitude of the fringes is actually so small that they do not show up even in the wavelength integrated map of $L_{tot}$ (see Fig.~\ref{fig3}). The fringes are presumably due to the glass window that protects the CCD chip inside the cryostat.} We applied some additional
corrections { in the lower part of the slit from 0$^{\prime\prime}$ to
  around 20$^{\prime\prime}$}, but point out that it could still be improved
by making a PCA analysis like in
\citet{marian+etal2008a,marian+etal2008b}. For the 2nd observation, residual
fringes are much weaker and can only be seen in Stokes $Q$. { The inversion
  code is not able to reproduce the pattern due to the combination of its limited number of degrees of freedom, the least-square minimization it uses, and the extent of the fringes in wavelength. The characteristic wavelength of the fringe pattern is around two times as wide as one of the solar spectral lines; with the chosen setup, the inversion code has no degrees of freedom (see the next paragraph) to produce such a pattern. In this case, two synthetic profiles will show small deviations: the one that includes the fringe pattern (which the inversion code cannot generate), and the one that ignores the additional modulation. We thus assume that the influence of the fringes on the
  results is negligible.} Comparing the two spectral lines at 1564.8 nm and 1565.2 nm, it is clear that most of the linear polarization signals seen in the more sensitive 1564.8 nm line are absent in the other{, like already noted in \citet{khomenko+etal2003}}; only the strongest linear polarization signals appear in both lines. Almost every profile along the slit shows a significant circular polarization signal in both spectra.

\begin{figure}
\centerline{\resizebox{8cm}{!}{\includegraphics{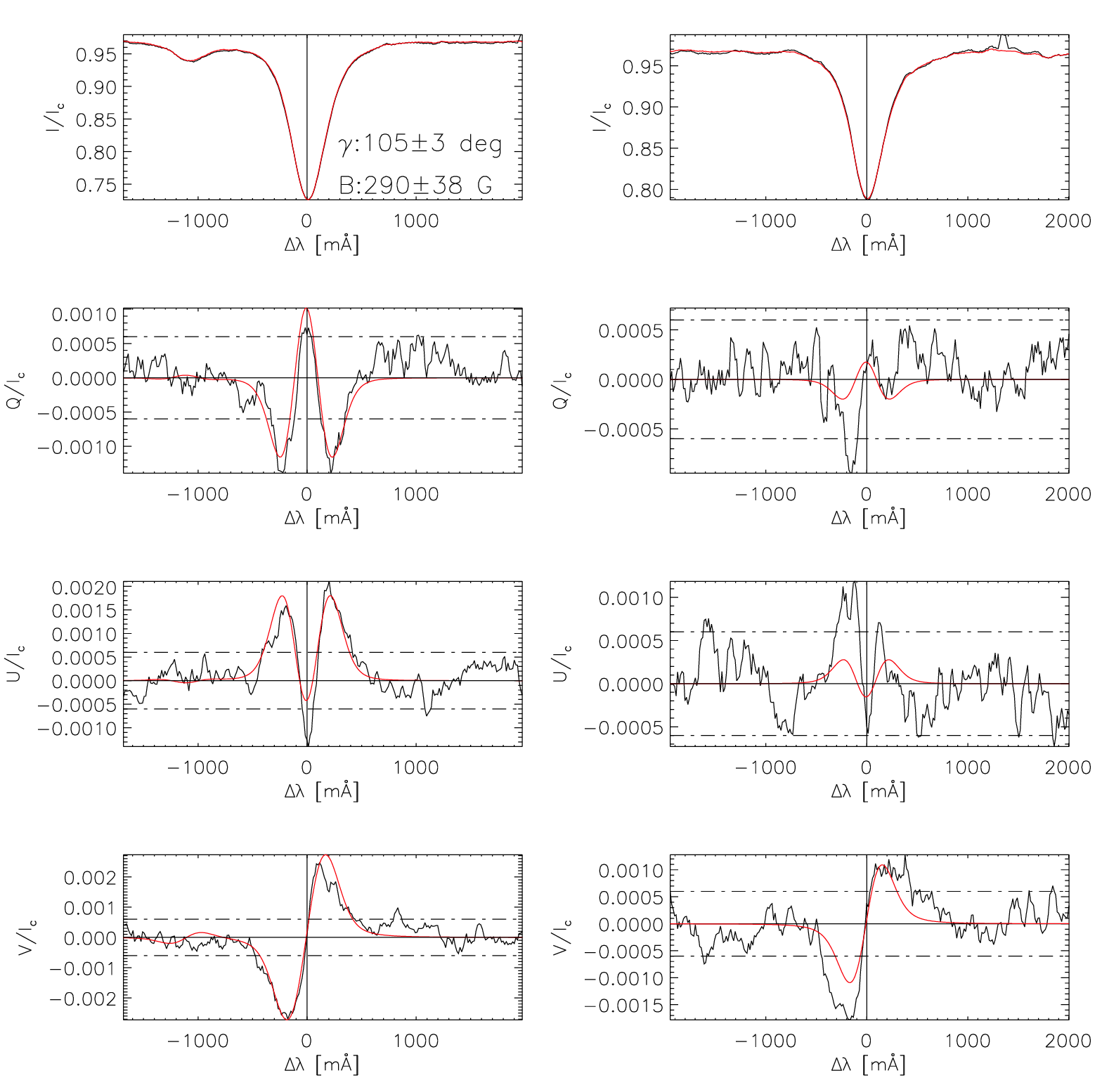}}}
\centerline{\resizebox{6cm}{!}{\includegraphics{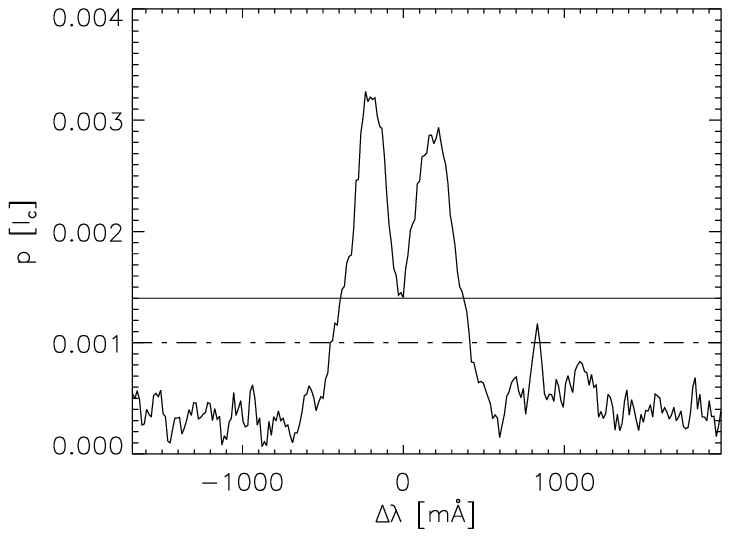}}}
\caption{{\em Top}: Example of inverted spectra of 1564.8 nm ({\em left
    column}) and 1565.2 nm ({\em right column}). {\em Black lines}: observed
  spectra of {\em top to bottom} $IQUV$; {\em red}: best-fit profile of the
  inversion. The {\em dash-dotted horizontal} lines mark three times the rms
  noise level, the {\em solid horizontal} line the zero level. The {\em
    vertical solid} line denotes the rest wavelength. {\em Bottom}:
  polarization degree  of 1564.8 nm with the threshold for inversion ({\em
    dash-dotted}) and final rejection ({\em horizontal solid
    line}). \label{fig10} }
\end{figure}

The noise level in the spectra is low enough that we could run a ``standard''
inversion of the spectra with the SIR code
\citep{cobo+toroiniesta1992}. ``Standard'' means that we have not modified the
inversion setup significantly to optimize it for the special type of data, to
get a hopefully unbiased first guess of the magnetic field properties without
enforcing a priori knowledge. We used the same setup as \citet{beck+etal2007}:
in each pixel we assumed one field-free and one magnetic atmosphere component,
with additional stray light. We only modified the inversion to use three nodes
in temperature for the magnetic component, and put equal weights for the
polarization components $Q,U$, and $V$. The parameters of the magnetic field
and the LOS velocities were taken to be constant with optical depth. The
2-C+stray light setup mimics observations of unresolved magnetic flux
(mag.~component) in equally unresolved intergranular lanes (field-free
component), whereas the stray light corresponds to the granular contribution
to the spectra\footnote{The average spectrum used as stray light contribution
  is blue-shifted and has unit { continuum} intensity.}. Figures \ref{fig10}, \ref{specs1}
and \ref{specs2} show several examples of observed spectra together with the
best-fit spectra retrieved by the inversion. We selected locations with a low
polarization degree and mostly larger linear than circular polarization signal
for the figures. For all these examples, the signal in the less sensitive
1565.2 nm line already is close to or below the noise level, whereas the
1564.8 nm line still yields a ``significant'' signal that the inversion code
can reproduce. 

For defining ``significant'', we used the polarization degree $p$. The routine that performed the inversion was using a threshold of $1\cdot 10^{-3}$ of $I_c$: if the maximal value of $p$ of either 1564.8 nm {\em or} 1565.2 nm was above this limit, the 2-C inversion with a magnetic component was done. As can be seen in Figs.~\ref{fig10} or \ref{p_exams} (right column, 3rd row), this level is reached already by (most probably) noise alone in two out of 13 cases for 1564.8 nm. We thus have set a slightly higher threshold for the analysis of the inversion result and rejected all profiles with $p<1.4\cdot 10^{-3}$ of $I_c$ in 1564.8 nm. In both maps, around 63 \% of the pixels are above the threshold that corresponds to 4 times the noise rms in $p$ ($3.4\cdot 10^{-4}$ of $I_c$). 

\begin{figure}
\centerline{\resizebox{7cm}{!}{\includegraphics{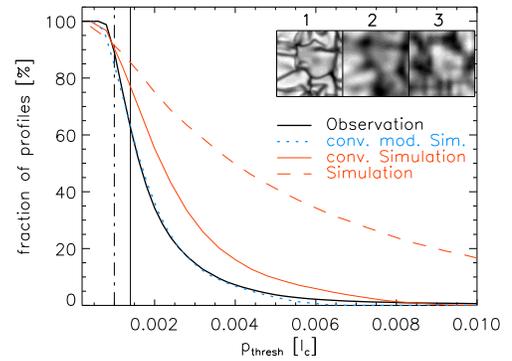}}}
\caption{Fraction of profiles with $p>p_{thresh}$. {\em Dash-dotted} and {\em solid vertical} lines denote the inversion and final rejection threshold in the observations, respectively. The inlets show continuum intensity maps of the original (1) and convolved simulation data (2), and a section of the same size from a TIP map (3).\label{fig_tresh}} 
\end{figure}

\citet{sheminova2008} found that about 70 \% of the surface area in a 2-D MHD simulation was covered with magnetic fields. OR07 give a fraction of 630 nm profiles of 87 \% and 35.5 \% for a polarization degree of 3 and 4.5 times the noise level of about $1\cdot 10^{-3}$ in their HINODE data. These numbers thus all agree that the area fraction covered by magnetic fields or showing polarization signal should be above 50 \%, but Fig.~\ref{fig_tresh} displays that the fraction of profiles with $p>p_{thresh}$ strongly depends on the threshold used. For comparison with a theoretical calculation, we derived the same fraction of profiles with $p>p_{thresh}$ from spectra of 1564.8 nm synthesized from a simulation run with the Co$^5$bold code described in \citet{schaffenberger+etal2005,schaffenberger+etal2006} and previously used in \citet{steiner+etal2008}. The snapshot was taken from the h20 run with an initial homogeneous horizontal magnetic field of 20 G. The simulation box corresponded to a 6\farcs6 square; its spatial resolution is much better than the present observations (cp.~inlets 1 and 3). We thus have convolved the simulations' spectra spatially with a Gaussian kernel of FWHM $\sim 0\farcs9$. We also added a noise level of 2.2$\cdot10^{-4}$ to them. The original simulation spectra without convolution and noise ({\em red dashed} in Fig.~\ref{fig_tresh}) have a much larger fraction of profiles with a high polarization degree than the observations. After the convolution, the curve already matches roughly that of the observations ({\em red solid}). We ascribe the remaining difference to the imperfect match of the spatial resolution, the different spatial sampling and the much larger solar area that enters in the observations (40$^{\prime\prime}$x70$^{\prime\prime}$). We only note that the convolved simulated spectra can be matched nearly perfectly to the observations by compressing the curve in the abscissa by a factor of 1.5 ({\em blue dotted line}). This compression can be understood as a reduction of the polarization degree that could be due to, e.g., the stray light that was not included in the convolution, or the usage of a Gaussian kernel with its steep drop with distance. The reduction by 1.5 may sound quite large, but one has to consider that the effect is to change, e.g., a 0.6 \% polarization level to 0.4 \% instead, for which only a small amount of additional stray light is needed. The slope of all curves changes clearly when reaching the noise level; the fraction of profiles is constantly close to 100 \% for thresholds below the noise. We suggest that the fraction of profiles above a polarization threshold can be used easily to match the spatial resolution of simulation spectra and observations, and also is a good indicator of the spatial resolution achieved in a specific observation.
\section{Inversion results\label{res}}
The inversion code constructs its synthetic spectra, $S$, by the following combination of inversion components:
\begin{equation}
S = \alpha \cdot S_{stray} +  (1-\alpha)\cdot( \beta \cdot S_{magnetic} +(1-\beta)\cdot S_{field-free}  ) \;,
\end{equation}
where $S_{stray}$ is the average profile of the whole map that is taken to
describe the stray light contribution and $\alpha$ its fractional contribution
to the spectrum. $\beta$ and $1-\beta$ are the relative contributions of the
magnetic and the field-free component. Assuming that $S_{stray}$ simply
corresponds to another field-free component with a fixed velocity, we define
the filling fraction $f$ of the magnetic fields by $f=(1-\alpha)\cdot
\beta$. The additional quantities from the inversion code are the field
strength, $B$, field inclination to the LOS, $\gamma$, field azimuth, $\psi$,
LOS velocities for each component, and the temperature stratification. The
observations were { on} disc centre, thus the LOS inclination $\gamma$ is identical to the field inclination to the surface normal. 
\begin{figure}
\centerline{\resizebox{6cm}{!}{\includegraphics{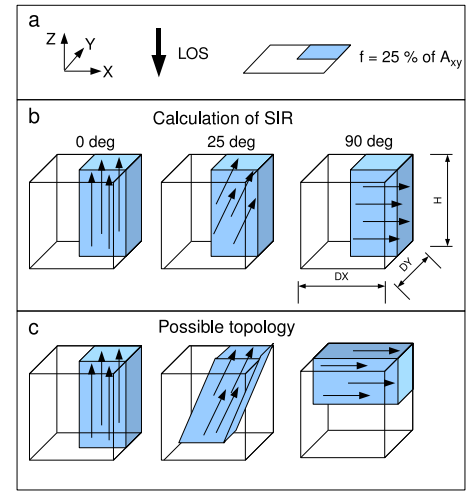}}}
\caption{{  Comparison of the geometry in the calculation of SIR and a possible topology inside the resolution element. {\em a, left to right}: reference frame, LOS direction, and filling fraction. {\em b}: geometry in the calculation by SIR for field inclinations to the LOS of {\em left to right} 0, 25, and 90 degrees. {\em c}: possible solution for real topology.}\label{fig_cube}}
\end{figure}

From the inversion results, we calculated an estimate of magnetic flux by 
\begin{equation}
\Phi = f\cdot B \cdot A \;, \label{flux_eq}
\end{equation}
 where $A$ is the area corresponding to a single pixel of 0\farcs5 x 0\farcs35.

For this estimate of the magnetic flux, we did not consider the field inclination. { We motivate this with the assumption that the volume contributing to the spectra from one pixel corresponds roughly to a (275 km)$^3$ cube. The spatial extent in the horizontal dimensions ($dx$, $dy$) is 360 km by 250 km. For the extension of the formation height, $H$, of the IR line at 1564.8 nm, \citet{cabrera+bellot+iniesta2005} give a range of around two dex in log $\tau$, e.g., from around 0 to -2 for velocity perturbations or from 0.5 to -2.5 for magnetic fields. In the HSRA model atmosphere \citep{gingerich+etal1971} they used, this range corresponds to around 250 km. Figure \ref{fig_cube} shows one possible solution how the calculation of SIR, which determines a filling fraction in the horizontal plane and then solves the radiative transfer along the LOS, could be related to the real topology inside the pixel. We assumed a constant filling fraction of 25 \% of the horizontal area $A_{xy}$ and plotted three cases for field inclinations to the LOS of 0, 25, and 90 degrees. To obtain the magnetic flux, the filling fraction has to be multiplied with the corresponding area perpendicular to the field direction. For horizontal magnetic fields, the respective area is, however, not $A_{xy}$ but $A_{xz}$ or $A_{yz}$, depending on the field azimuth. If $dx\sim dy\sim H$, the surfaces of the cube are equal and thus the field orientation can be ignored to first order. This is only approximative, as for example in the 90-deg-case, SIR calculates the synthetic profile assuming that the horizontal fields are present everywhere inside the formation height, which not necessarily will give the same result as if they fill only half the formation height with doubled filling fraction.}
\begin{figure}$ $\\$ $\\
\centerline{\resizebox{7.5cm}{!}{\includegraphics{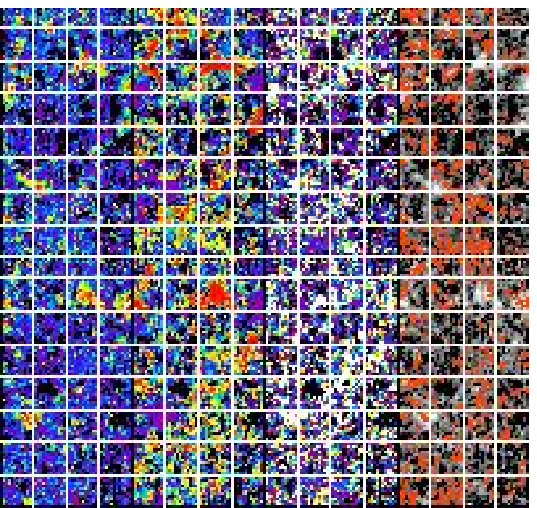}}}$ $\\$ $\\
\centerline{\resizebox{3.5cm}{!}{\includegraphics{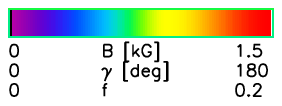}}\resizebox{3.5cm}{!}{\includegraphics{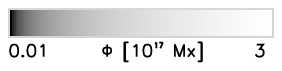}}}
\caption{Inversion results of Op.~001 ({\em top row}) and Op.~005 ({\em bottom
    row}). {\em Left to right}: field strength, LOS inclination, filling
  fraction, total flux. {\em Red contours} in $\Phi$ trace strong linear
  polarization signals. { The display ranges are given by the labels of the
    color bar at bottom. Pixels not inverted are shown {\em black}.}\label{fig4}}
\end{figure}

Figure \ref{fig4} shows maps of $B$, $\gamma$, $f$ and $\Phi$ for the two observations. Magnetic fields with kG or more are restricted to vertical fields ({\em 1st and 2nd column}, $\gamma\sim 0$ or 180$^\circ$). The kG fields show a high magnetic filling fraction, but also parts with horizontal fields can reach the same level in $f$. The more inclined fields, however, do not appear prominently in the flux map ({\em rightmost column}). We overplotted contour lines tracing the strongest linear polarization signal (cf.~Fig.~\ref{fig3}) onto the flux map. Locations of large magnetic flux are almost exclusively related to vertical fields with little linear polarization signal. 

The total amount of magnetic flux in both maps calculated with Eq.~(\ref{flux_eq}) is around 6$\cdot 10^{20}$ Mx. There is a slight imbalance of 53 to 47 \% for magnetic flux parallel or antiparallel to the LOS. Including the field inclination in the calculation to obtain the LOS magnetic flux by multiplying $\Phi$ with cos $\gamma$, the ratio is reversed to 47 to 53 \%. As the fraction is close to 50 \%, we think that the FOV gives a sufficiently large statistical sample of quiet Sun regions where no strong imbalance in favor of a special field orientation is to be expected. The histogram of inclinations using the full range from 0 to 180 deg was also fairly symmetric, so we have in the following converted field inclinations above 90 degrees to their corresponding values in the range from 0 to 90 degrees. 
\begin{figure}
\centerline{\resizebox{7.5cm}{!}{\includegraphics{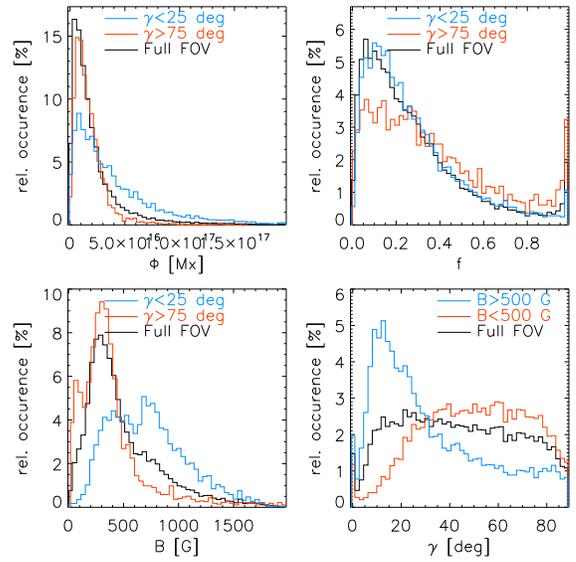}}}
\caption{Histograms of ({\em clockwise, starting left top}) magnetic flux, filling fraction, field inclination, field strength. {\em Blue} and {\em red lines} correspond to the least/most inclined fields, {\em black lines} to the full FOV. \label{fig5}}
\end{figure}

Figure \ref{fig5} shows histograms of $B$, $\Phi$, $\gamma$ and $f$. We calculated the histograms for three different samples there: the full FOV ({\em black}), all locations with nearly vertical magnetic fields ({\em blue}, $\gamma<25^\circ$), and nearly horizontal magnetic fields ({\em red}, $\gamma>75^\circ$). For the histogram of $\gamma$ we separated into magnetic fields above and below 500 G instead. The visual impression of the 2-D maps is confirmed by the histograms: the nearly vertical fields correspond to the locations with large magnetic fluxes and high field strength (300-1000 G); the nearly horizontal fields are limited to below around 500 G. 

The distribution of magnetic field strength for the full FOV agrees with previous observations of the IR lines \citep{khomenko+etal2003,marian+etal2008a} or the field strength distribution derived by \citet{sheminova2008} from 2-D MHD simulations, especially in the point of the most probable field strength of around 250 G. We point out that a) the ``bump'' at around 1.5 kG that commonly is found from the analysis of 630 nm data \citep[e.g.,][ OR07]{cerdena+etal2006} is missing, and that b) the decrease of relative occurrence for fields below 250 G is not due to the polarization detection limit. Figure \ref{specs2} shows 3 examples of weak fields below 200 G (2nd row) that still could be reliably reproduced by the inversion. The spatial resolution of observations is also crucial for the detection of these weak fields (see Fig.~\ref{fig_tresh}). The distribution of the nearly horizontal fields ({\em red curve at bottom left} of Fig.~\ref{fig5}) shows an enhanced fraction of these weak fields below 200 G which implies that the detection limit due to the noise is not yet reached at 200 G. In the question of the seemingly preference for kG fields in the analysis of 630 nm data, \citet{marian+etal2006} have cautioned that the thermodynamics have to be treated consistently for the 630 nm line pair in the weak field limit. \citet{rezaei+etal2007} were able to derive a distribution without the bump using a SIR inversion of 630 nm data of lower spatial resolution than the nowadays available HINODE observations. It would be worthwhile to analyze the HINODE data without employing the Milne-Eddington (ME) approximation to see if the field strength distribution changes with a more sophisticated treatment of the thermodynamics. 

The field inclination shows some clear trends: for fields below 500 G, an inclination above 45$^\circ$ is preferred with a plateau of constant probability up to 90$^\circ$, whereas strong fields above 500 G are restricted to an inclination of 35$^\circ$ or less. The distribution for the full FOV shows a small peak near 0$^\circ$ inclination ($\equiv$ the very small area fraction of ``network'' fields inside the FOV), and then a rapid increase of occurrence up to 15$^\circ$ followed by a slow reduction towards 90$^\circ$. The shape of the distribution matches more closely to the one given by OR07 in their Fig.~3 for the MHD simulation run of \citet{voegler+etal2005} than the one these authors derived from their observations. 
\begin{figure}
\centerline{\resizebox{7.5cm}{!}{\includegraphics{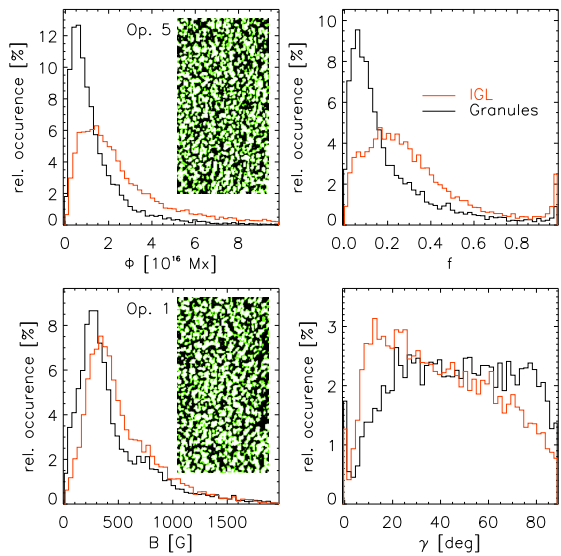}}}
\caption{The same histograms as in Fig.~\ref{fig5} but separated into granules
  ({\em black lines}) and intergranular lanes ({\em red}). { The inlets
    show the masks for the separation.}\label{fig9}}
\end{figure}

A second method of classification into different type of structures is to
separate the FOV into granules and intergranular lanes (IGLs). We derived a
mask using the continuum intensity maps and a threshold of 0.995 and 1.005 of
$I_c$ to define IGLs { ({\em black} in the inlets in Fig.~\ref{fig9})} and
granules { ({\em white})}, respectively. Locations with an intensity
between the two values were excluded { ({\em green})}. { The masks
  retain the granulation pattern of the continuum intensity. At the
  near-infrared wavelength of the observations and at their spatial sampling,
  magnetic elements do not show up as bright in the continuum on disc center,
  and thus should not incidentally have been counted as part of the granular sample.} We then
calculated the histograms of the same quantities as in Fig.~\ref{fig5}. The
pattern is very similar to that in Fig.~\ref{fig5}: in the IGLs, one has a
predominance of more vertical, stronger fields with larger magnetic flux. A
strong difference between granules and IGLs is seen in the magnetic filling
fraction. IGLs show a larger contribution of high filling fractions centered
at 0.2-0.3, whereas the granules show a distribution that peaks close to $f=0$
and drops steeply for larger $f$. This indicates that the magnetic flux in (or
above) granules is more diffuse  than in the IGLs and gets rather dispersed
than concentrated, whereas in the IGLs the converging granular flows advect
magnetic flux and concentrate it.
\begin{figure}
\centerline{\resizebox{5cm}{!}{\includegraphics{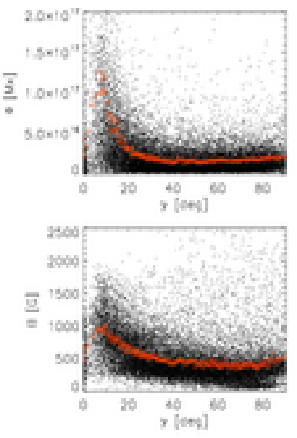}}}
\centerline{\resizebox{5cm}{!}{\includegraphics{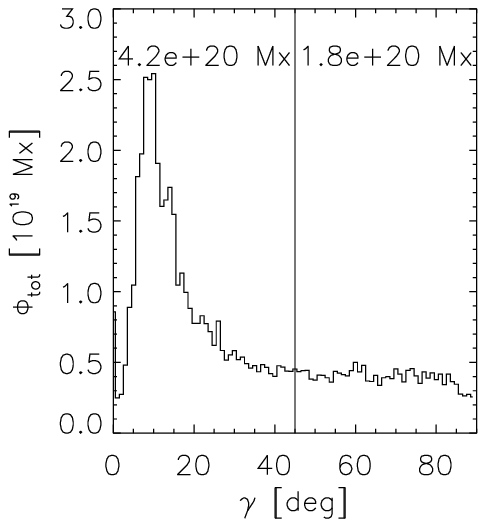}}}
\caption{Scatterplots of magnetic flux vs inclination ({\em top}) and field strength vs inclination  ({\em middle}). {\em Red crosses} are binned data. {\em Bottom}: total magnetic flux in the full FOV as function of field inclination in 1 deg bins. The {\em black vertical line} marks 45$^\circ$; the numbers give the total magnetic flux of all fields below or above this limit.\label{fig6}}
\end{figure}

The relative amount of magnetic flux in concentrated vertical fields in IGLs and the diffuse weak horizontal fields in granules is of strong interest, e.g., for reconciling Hanle and Zeeman measurements with low spatial resolution that yielded quite different average magnetic flux densities, with Zeeman-based observations lower by around an order of magnitude. The histograms of Figs.~\ref{fig5} and \ref{fig9} indicated a tight relation between field strength, magnetic flux and field inclination. We have thus also created scatterplots of inclination vs the other two quantities ({\em top two panels} of Fig.~\ref{fig6}). We remark that the field inclination did {\em not} enter into the calculation of the magnetic flux, but only filling fraction and field strength. In the scatterplots, there is a clear trend that the inclined fields above 45$^\circ$ are limited to below 500 G and below around 5$\cdot10^{16}$ Mx. Larger fluxes are only achieved by nearly vertical fields \citep[cp.][their Fig.~6]{ishikawa+tsuneta2009}. To obtain the total amount of magnetic flux as function of inclination, we added the magnetic fluxes of all locations with a field inclination in $[\gamma,\gamma+1^\circ]$ ({\em lower panel}). The distribution is similar to the scatterplot at top; we find that around 2/3 of the total magnetic flux comes from magnetic fields with $\gamma<45^\circ$.

\begin{figure}
$ $\\
\centerline{\resizebox{6cm}{!}{\hspace*{1cm}\includegraphics{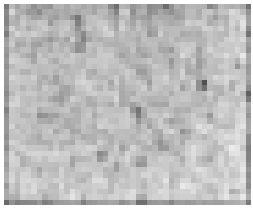}}}$ $\\
\caption{Linear polarization map $L_{tot}$ of an observation of quiet Sun with TIP@1.56 micron from 17.8.2003. Tick marks are in arcsec. \label{old_tip}}

\end{figure}

In observations with low spatial resolution, the weaker and more inclined magnetic fields could not be detected because their linear polarization signal was spatially smeared out to values below the noise level \citep[see Fig.~\ref{old_tip}, observations are described in detail in][]{khomenko+etal2005,marian+etal2008a}. These data of 2003 were taken in the same spectral range as the present investigation, with an identical integration time of 30 sec per spectrum, but an image correction with only a correlation tracker to remove image shifts during the exposure. A single position at (48$^{\prime\prime}$,22$^{\prime\prime}$) in a 60$^{\prime\prime}$x35$^{\prime\prime}$ FOV shows significant linear polarization signal that would indicate inclined magnetic fields. { \citet{khomenko+etal2003}, however, were able to detect linear polarization signals in 50-sec integrated data from TIP taken on July 29th 2000 and September 5th 2000\footnote{See http://www3.kis.uni-freiburg.de/$\sim$cbeck/TIP\_archive/\\TIP\_archivemain.html}. They found that around 20 \% of the pixels inside the FOV showed a significant linear polarization signal that could be analyzed quantitatively.  }
\section{Discussion\label{disc}}
\paragraph{{ General properties of the QS magnetic fields}}
The weak horizontal magnetic field of the solar internetwork regions is only accessible to observations of high spatial resolution and high polarimetric sensitivity. The presence of these fields has been expected from Hanle measurements, and traces of them have been found in some observations before \citep{lites+etal1996,khomenko+etal2003,harvey+etal2007,marian+etal2008,marian+etal2008a}. They have only recently been seen { prominently} with data from the spectropolarimeter on-board the HINODE satellite \citep{kosugi+etal2007}, as reported by OR07 and LI08. From a comparison of the present Fig.~\ref{fig3} with Figs.~2 and 7 of LI08 we conclude that the present observations trace the same solar structures, i.e., ground-based observations aided by real-time adaptive optics correction can reach the necessary spatial resolution during spells of excellent seeing conditions, and hence, the results of the analysis should be compatible. 

Some of the present findings are at odds to the HINODE data, but we think that
we can trace back most of these contradictions to the analysis methods, the
different magnetic sensitivity of the spectral lines, and finally, to a
physical reason. The distributions of field strength show again the mismatch
between IR and VIS found in previous studies: a significant fraction of kG
fields in VIS (OR07) that do not appear for IR
data. \citet{bellotrubio+collados2003}, \citet{marian+etal2006}, and
\citet{rezaei+etal2007} have shown that this feature is related to the noise
level of the data and the thermodynamics, because IR and VIS spectra together
can be reproduced simultaneously with weak fields
\citep{marian+etal2008a}. The usage of a ME method in OR07 possibly has also a
contribution to this effect. The field strength distribution here peaks at
around 250 G, whereas \citep{orozco+etal2007} give 90 G. This can again be
related to the weak-field limit of the 630 nm lines together with the ME
approximation that makes the VIS lines less reliable than the stronger
splitting IR lines. The magnetic filling fraction $f$ compares fairly well to
one minus the stray light contribution of OR07; magnetic filling fractions
peak at around 0.1 here and at 0.2 in OR07. The field inclinations differ
most: we do not find an increase in the distribution  for horizontal fields,
but a { probability declining slowly with inclination for fields with
  inclinations above 20$^\circ$}. OR07 and LI 08 both found a clear dominance of horizontal over vertical fields and transversal over longitudinal flux, respectively. With the same caveat than in LI08 that the vertical and horizontal extent of the transverse fields actually cannot be derived at once from the spectra, we found that 2/3 of the total flux comes from fields with inclinations below 45$^\circ$, whereas LI08 give a ratio of 5 {\em in favor} of the transversal fields.
\begin{figure*}
\sidecaption
\resizebox{11.5cm}{!}{\includegraphics{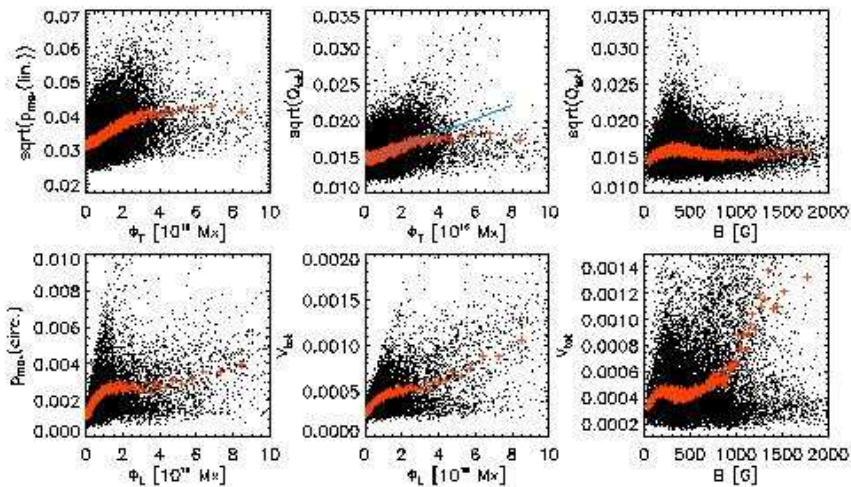}}
\caption{Scatterplots of transversal and longitudinal flux vs maximal linear ({\em  left top}) and circular polarization degree ({\em left bottom}), same vs total linear ({\em middle top}) and circular polarization ({\em middle bottom}), and scatterplot of field strength vs total linear ({\em right top}) and circular polarization ({\em right bottom}). {\em Black}: all data points; {\em red}: after binning. The {\em blue line} in the {\em middle top} panel is a linear fit to the binned data. \label{fig7}}
\end{figure*}
\paragraph{Reliability of proxies in magnetic flux determination}
As we were able to successfully run an inversion on most of the pixels due to the low noise level of our data, we have the information on field strength, field inclination and magnetic flux with some certainty. This allows to test the validity of some of the approximations in LI08 who convert the total linear and circular polarization signal into their estimates of transversal and longitudinal flux. To this extent, we calculated the longitudinal ($\Phi_L$) and transversal  ($\Phi_T$) flux from the inversion results by 
\begin{eqnarray}
&&\Phi_L = f \cdot B \cdot \cos \gamma \cdot A \label{flux_eq1}\\\nonumber
&&\Phi_T = f \cdot B \cdot \sin \gamma \cdot A \;. 
\end{eqnarray}
As first order approximation, the  circular polarized Stokes $V$ signal should scale with $\Phi_L$, whereas the linear polarization $L=\sqrt{Q^2+U^2}$ should scale with $\Phi_T^2$ \citep{jefferies+etal1989}. From the observed profiles, we derived the maximal linear and circular polarization of the 1564.8 nm line ($L_{max}$ and $V_{max}$, see Sect.~\ref{obs}) and the total linear and circular polarization ($L_{tot}$ and $V_{tot}$, see Sect.~\ref{ana}). In the derivation of $L_{tot}$, the noise in both $Q$ and $U$ enters. Following LI08, this contribution can be minimized by rotating the linear polarization signal to the ``preferred azimuth frame'' where the signal is concentrated in Stokes $Q$ (see Appendix \ref{lin_cal}). We will thus in the following use $Q_{tot}({\rm rotated}) = \int |Q(\lambda)|d\lambda$ after the rotation of the spectra as a measure of the linear polarization instead of $L_{tot}$. Figure \ref{fig7} shows scatterplots of $\Phi_T$ vs $\sqrt{Q_{max}}$ and $\sqrt{Q_{tot}}$, and  $\Phi_L$ vs $V_{max}$ and $V_{tot}$. We also added a plot of field strength vs $\sqrt{Q_{tot}}$ and  $V_{tot}$. It can be seen that there is {\em on average} a clear relation between $\Phi_T$ or  $\Phi_L$ and the corresponding quantities derived directly from the profiles ({\em red points}), but the scatter around this relation is large (around 30 \% difference between mean and minimum or maximum values). 

The polarization  amplitudes, and hence, also the integrated polarization amplitudes measure the ratio of polarized to unpolarized light. Their absolute values strongly depend on the thermodynamical structure of the atmosphere and the resolution of structures. This gets even worse for VIS lines in the weak-field limit where the field strength also changes the amplitudes of polarization lobes, not only their location in wavelength. In Appendix \ref{lin_cal} we tested the derivation of calibration curves between transversal magnetic flux and $Q_{tot}$ under varying assumptions (constant inclination, constant  flux, temperature variation). We find that only varying the temperature stratification at constant magnetic flux is almost sufficient for producing the scatter seen in Fig.~\ref{fig7}. In total, the scatter of the relationship and the different calibration curves that can be obtained by, e.g., only varying the inclination by 10 degrees put the derivation of magnetic flux from proxies as $Q_{tot}$ or $V_{tot}$ into some doubt.

However, besides from the observations with HINODE that favor a dominant
fraction of horizontal fields recently also some simulation results have
suggested more horizontal magnetic fields then
previously. \citet{schuessler+voegler2008} and \citet[][ST08]{steiner+etal2008} have reported the ratio of horizontal to vertical
fields in their respective simulations and give numbers from 2 to 6, in
accordance with the factor of 5 derived by LI08. This seems incompatible with
the present investigation, but both simulations show that the ratio is
strongly dependent on height in the atmosphere. ST08 show in their Fig.~1 the
field strengths as function of height for two simulation runs (v10=initial 10
G vertical field, h20= 20 G horizontal). At a height of around 200 km, which
is appropriate for the IR lines at 1565 nm used here, the ratio of v10 is
actually in favor of the longitudinal fields, $B_{hor}/B_{ver} \sim
0.82$. This is significantly closer to the ratio of $\Phi_{\gamma >
  45^\circ}/\Phi_{\gamma < 45^\circ} = 0.42$ than a factor of 5 that ST08
derive after synthesizing 630 nm spectra of their simulation and analyzing
them in the same way as LI08. 
\paragraph{Vertical unsigned flux density} Various authors have derived the
average unsigned flux density of the vertical magnetic fields from Zeeman
sensitive spectral lines
\citep{keller+etal1994,lites2002,cerdena+etal2003,khomenko+etal2005,cerdena+etal2006,cerdena+etal2006a,orozco+etal2007,marian+etal2008a,lites+etal2008,carroll+kopf2008}.
In the older literature, magnetic field proxies like the Stokes $V$ amplitude
or integrated area were used; in the most recent ones an inversion of
spectra. The obtained values for $\langle B_z\rangle$ scatter strongly, depending on the observations used or the analysis method employed, but in generally yield values between 5 to 30 G. LI08 in addition give a value of around 50 G for the transversal magnetic flux. Observations using the Hanle effect yielded significantly larger values of around 100 G \citep[e.g.,][]{trujillobueno+etal2004}.

\begin{figure}
\resizebox{7.5cm}{!}{\includegraphics{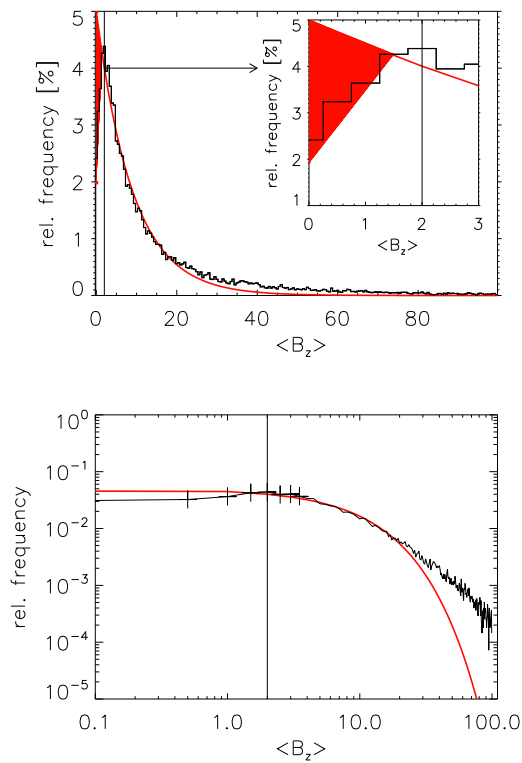}}
\caption{Histograms of vertical flux density in linear ({\em top}) and
  logarithmic scale ({\em bottom}). The {\em vertical lines} mark 2 G. { {\em Black} and {\em red} lines show the observations and an exponential decay law. The inlet in the upper panel shows a magnification of the low flux range.}\label{vert_b}}
\end{figure}

Figure \ref{vert_b} shows the distribution of $\langle B_z\rangle$ in our observations, derived from the longitudinal flux by division with the area $A$
(cf.~Eq.~(\ref{flux_eq})); the bin size was 0.5 G. The distribution shows a
maximum near 2 G, but otherwise is in good agreement with a
monotonically increasing probability for weaker fluxes ({\em solid red line},
exponential decay law). The difference between the exponential decay and the
observed distribution for $\Phi<$ 2 G actually only covers a small area of far
less than 10 \% relative frequency ({\em red filled area} { in the upper panel of Fig.~\ref{vert_b} from 0 to 2 G, see the inlet}). In the
double-logarithmic display in the {\em lower half}, we displayed the first 7
histogram points with crosses. It can be clearly seen that from high fluxes up
to the maximum the points follow well the exponential law; only the first
four points referring to the weakest fluxes below 2 G deviate. Taking into
account that the noise imposes a { {\em hard}} threshold for the detection of the smallest magnetic fluxes, we believe that the occurrence of the maximum is
fully artificial. Very weak fluxes produce smaller and smaller polarization
signal that can escape detection completely or are eventually rejected by the inversion threshold. { \citet{graham+etal2009} claim that the distribution of the magnetic flux density cannot be determined from observations because of the peaked appearance of the probability distributions like in Fig.~\ref{vert_b}. However, given the actually small deviation between the observed distribution and an exponential decay law predicted by MHD simulations,}
we think that the probability distribution function of magnetic flux can
be derived from observations { when one takes into account that the weakest fluxes {\em must} escape detection due to the noise level.

Using the strongly Zeeman-sensitive IR lines at 1.56 $\mu$m, it is possible to disentangle the field strength $B$ on the part of the surface that actually is covered by fields from the area filling fraction $f$ of the almost certainly
unresolved magnetic structure. This makes an important difference for the calculation of the magnetic flux density, compared with the spectral lines at 630 nm used by the HINODE spectropolarimeter and in the work of \citet{graham+etal2009}. The 630 nm lines are in the weak field limit for the weak magnetic fields of the quiet Sun, i.e., they suffer from a strong interplay between thermodynamics and magnetic field properties, as also demonstrated in Appendix \ref{lin_cal}. It is more difficult to determine $B$ and $f$ accurately for these lines, even if one takes care about the thermodynamics \citep{marian+etal2006}.}

{ Normalizing the observed magnetic flux to the complete FOV, including the locations with no  significant polarization signal, the average values for the total, longitudinal and transversal flux using Eqs.~(\ref{flux_eq}) and (\ref{flux_eq1}) are 22\,G, 16\,G, 
and 11\,G, respectively. If one considers only the area with actually observed
polarization signal, the corresponding values are 26\,G, 20\,G, and 13\,G. 
The values are on the upper range of Zeeman based measurements. As mentioned by, e.g., \citet{graham+etal2009}, the cancellation of Stokes $V$ signals due 
to unresolved opposite polarities inside a resolution element causes underestimation 
of the real amount of magnetic flux. 
Our observations with a 0\farcs5$\times$0\farcs37 sampling are
prone to this. The average flux densities thus can only be taken as lower
limit for the total magnetic flux, but we believe that they actually give a
rather solid limit for the total magnetic flux observed at the 1$^{\prime\prime}$ resolution achieved.}
\vfill
\section{Summary \& conclusions\label{concl}}
We have analyzed two long-integrated data sets of quiet Sun regions obtained
under extremely good seeing conditions with the TIP at 1.56 $\mu$m. We find
that the data allow to study the weak horizontal internetwork fields that
appeared prominently in HINODE observations recently (OR07 and LI08). Thanks
to a lower noise level { than in the HINODE data} and the higher sensitivity of the IR lines, we could
run an inversion of the maps with SIR that saves us from using strong
approximations in the derivation of magnetic field properties. We find that
the more inclined/``horizontal'' fields ($\gamma>45^\circ$) are generally
weaker ($<$500 G) and contain in total {\em less} magnetic flux than the less
inclined/``vertical'' fields. As the data properties (spatial scales of
observed structures, pixels with significant polarization signal) and the
inversion results (magnetic filling fraction, relation to granules/IGLs) seem
roughly to match to the HINODE data and their analysis in OR07 and LI08, we
ascribe the { reversal} of the ratio of transversal to longitudinal fields (HINODE data: 5, IR data: 0.42) to a) the different formation height of the spectral lines and b) the analysis methods (ME inversion, calibration from integrated polarization signal to magnetic flux). We conclude that the issue should best be investigated with simultaneous IR/VIS data, as simulations predict a strong dependence of the ratio on geometrical height \citep{steiner+etal2008}. { Whereas the existence of a significant fraction of inclined magnetic fields is confirmed by the present IR observations, their influence on the dynamics of the solar atmosphere is not clear yet. They do not seem to have an impact on the granulation pattern, but could contribute to chromospheric heating as discussed in \citet{ishikawa+tsuneta2009}.} This issue can be addressed by a combination of repeated spectropolarimetry of photospheric fields with chromospheric diagnostics.
\begin{acknowledgements}
The VTT is operated by the Kiepenheuer-Institut f\"ur Sonnenphysik (KIS) at the
Spanish Observatorio del Teide of the Instituto de Astrof\'{\i}sica de Canarias (IAC). R.R.~acknowledges support by the Deutsche Forschungsgemeinschaft under grant SCHM 1168/8-1. C.B.~thanks M.Collados for being the only co-observer to show up during the campaign. { We thank O.~Steiner (KIS) for making  the CO$^5$BOLD simulation data available to us.}
\end{acknowledgements}
\bibliographystyle{aa}
\bibliography{references_luis_mod}
\clearpage
\begin{appendix}
\section{Examples of inverted spectra \label{appa}}
\begin{figure*}
\fbox{\resizebox{6cm}{!}{\includegraphics{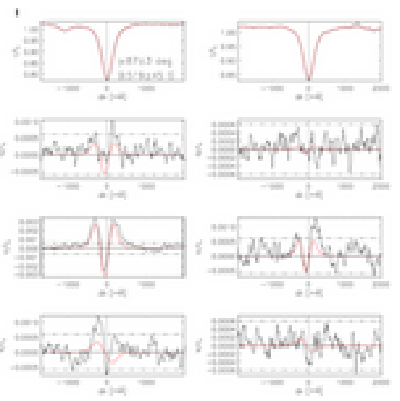}}}\fbox{\resizebox{6cm}{!}{\includegraphics{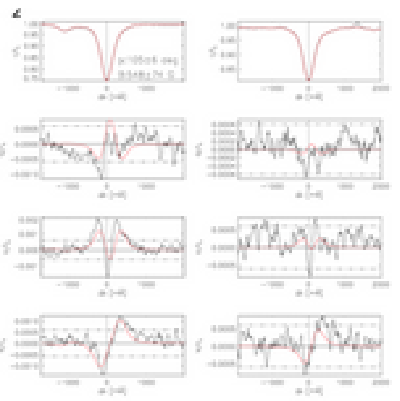}}}\fbox{\resizebox{6cm}{!}{\includegraphics{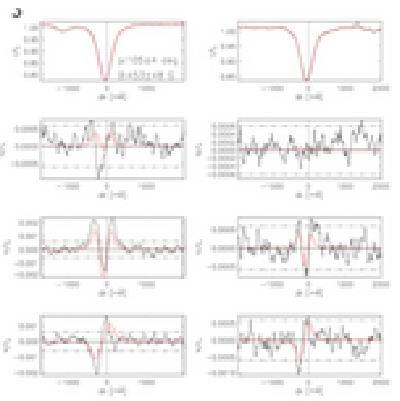}}}\\
\fbox{\resizebox{6cm}{!}{\includegraphics{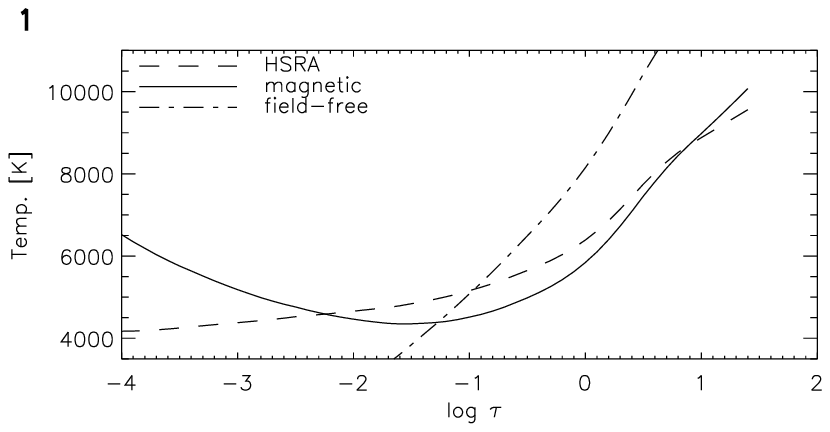}}}\fbox{\resizebox{6cm}{!}{\includegraphics{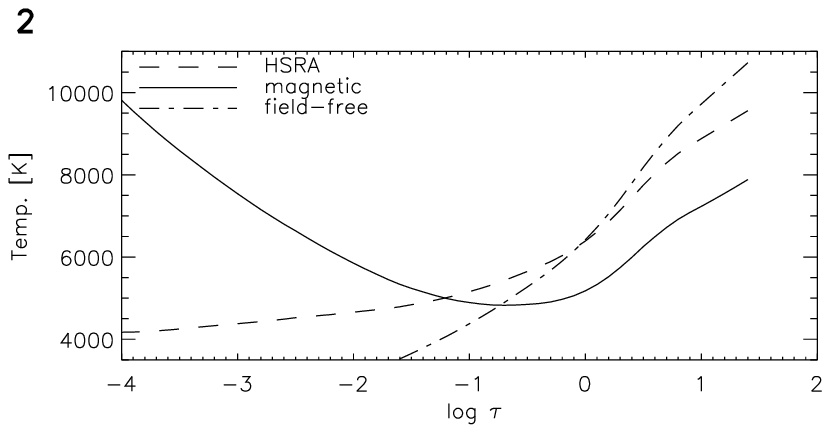}}}\fbox{\resizebox{6cm}{!}{\includegraphics{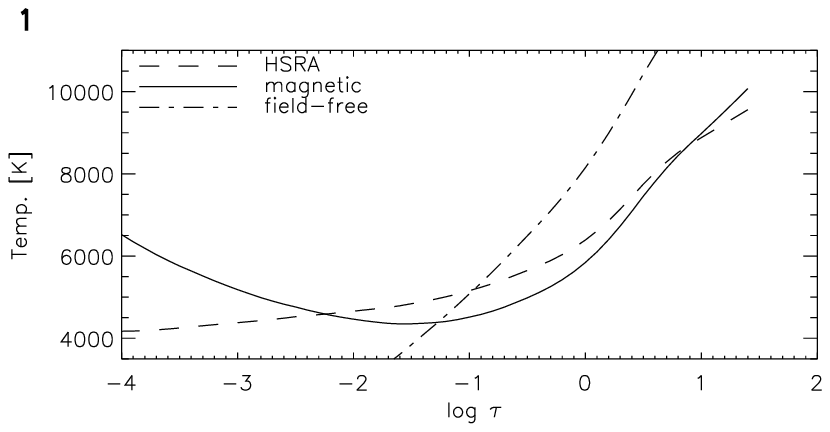}}}\\
\fbox{\resizebox{6cm}{!}{\includegraphics{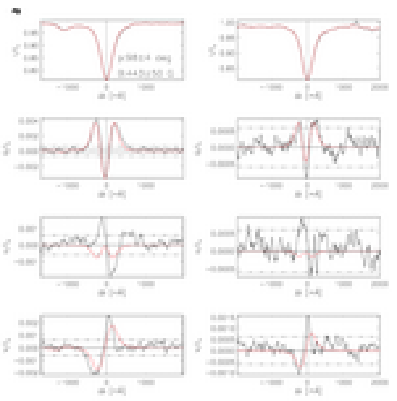}}}\fbox{\resizebox{6cm}{!}{\includegraphics{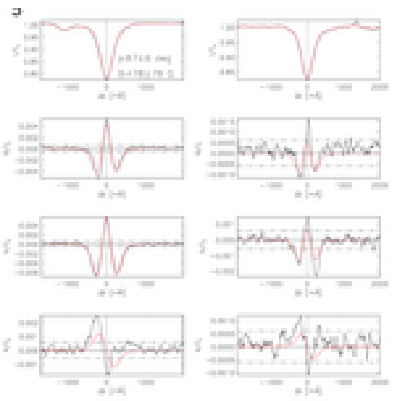}}}\fbox{\resizebox{6cm}{!}{\includegraphics{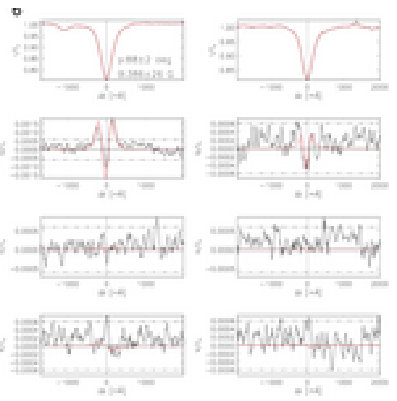}}}\\
\fbox{\resizebox{6cm}{!}{\includegraphics{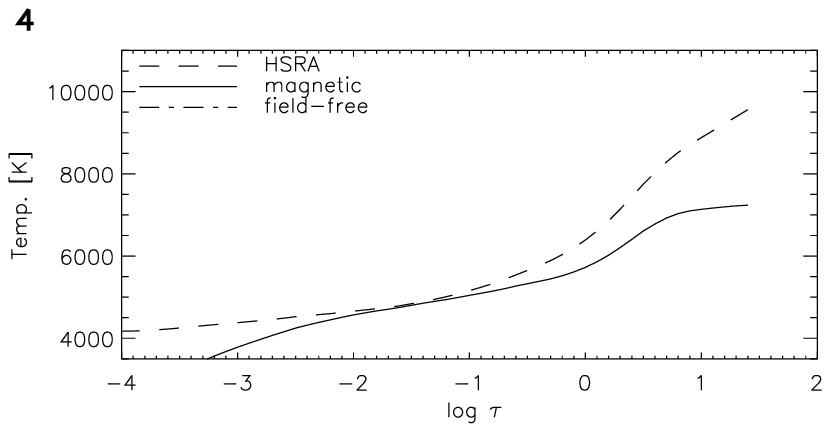}}}\fbox{\resizebox{6cm}{!}{\includegraphics{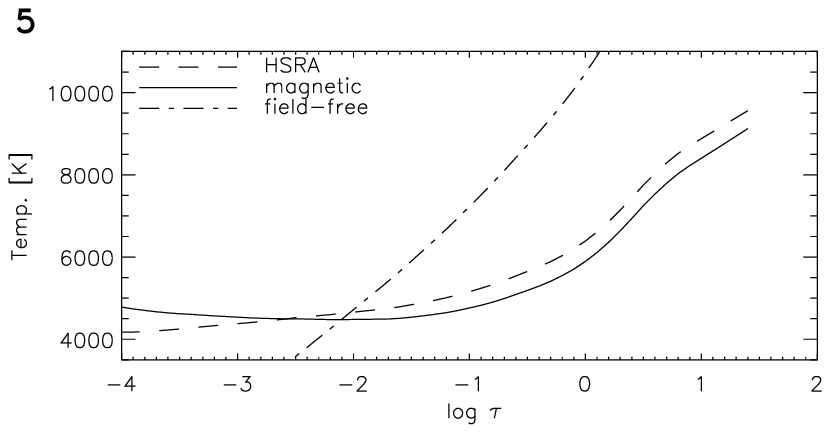}}}\fbox{\resizebox{6cm}{!}{\includegraphics{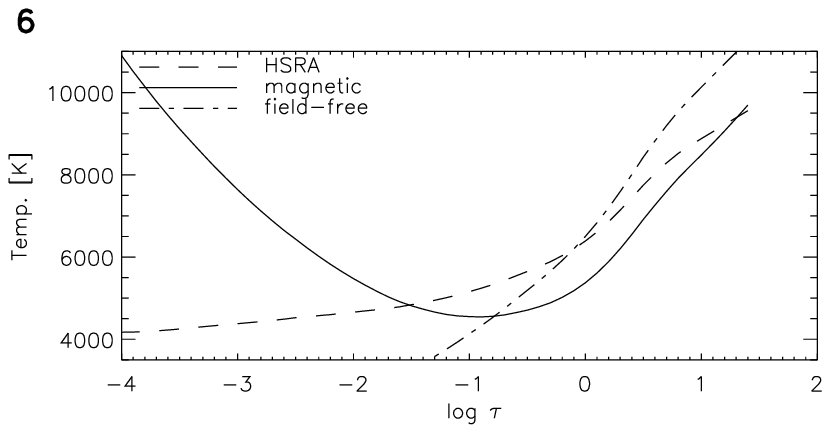}}}
\caption{Examples of observed ({\em black line}) and best-fit spectra ({\em red}) in Op.~001 ({\em { 1st and 3rd row}}). The {\em dash-dotted horizontal} lines in $QUV$ mark three times the rms noise level, the {\em solid horizontal} line the zero level. The {\em vertical solid} line denotes the rest wavelength. The {\em { 2nd and 4th row}} show the corresponding temperature stratifications of the magnetic component ({\em solid}), the field-free component ({\em dash-dotted}), and the HSRA atmosphere that is used as initial model ({\em dashed}). Field strength and LOS inclination { and their respective errors} are given in the plot of Stokes $I$ of 1564.8 nm ({\em upper left in each panel}); { the number of each profile is given in the {\em upper left} corner of each panel}.\label{specs1}}
\end{figure*}

\begin{figure*}
\fbox{\resizebox{6cm}{!}{\includegraphics{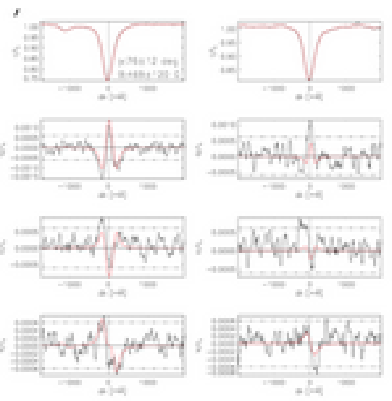}}}\fbox{\resizebox{6cm}{!}{\includegraphics{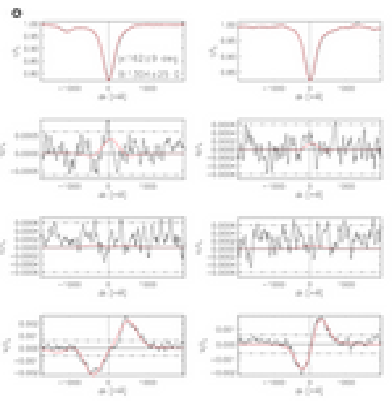}}}\fbox{\resizebox{6cm}{!}{\includegraphics{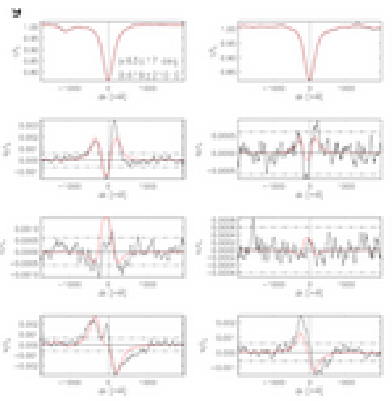}}}\\
\fbox{\resizebox{6cm}{!}{\includegraphics{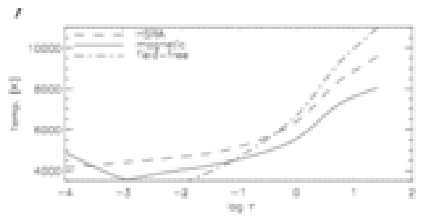}}}\fbox{\resizebox{6cm}{!}{\includegraphics{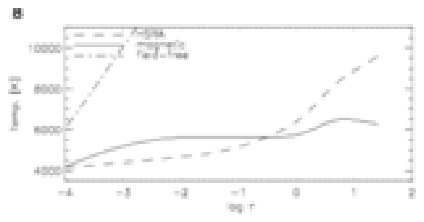}}}\fbox{\resizebox{6cm}{!}{\includegraphics{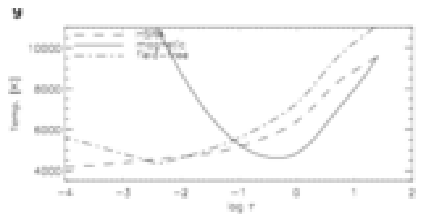}}}\\
\fbox{\resizebox{6cm}{!}{\includegraphics{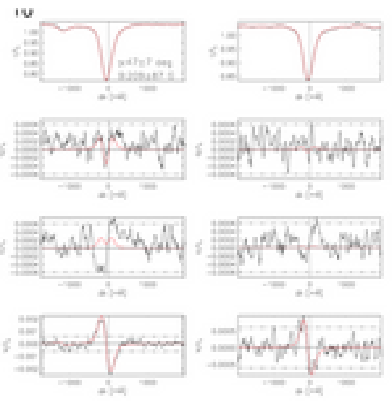}}}\fbox{\resizebox{6cm}{!}{\includegraphics{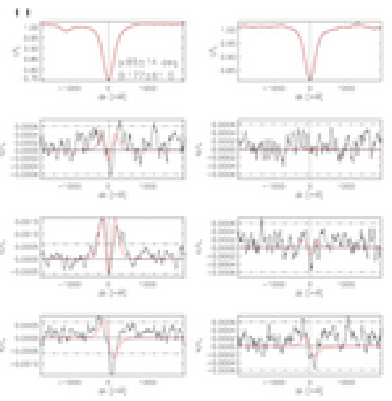}}}\fbox{\resizebox{6cm}{!}{\includegraphics{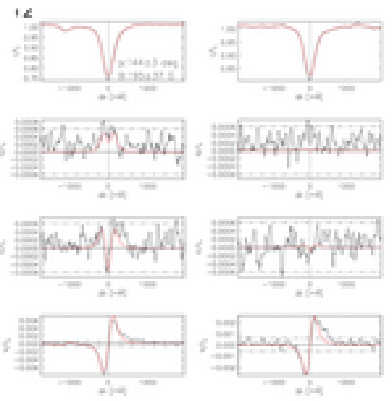}}}\\
\fbox{\resizebox{6cm}{!}{\includegraphics{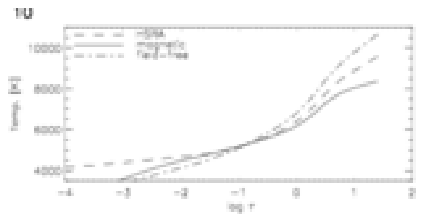}}}\fbox{\resizebox{6cm}{!}{\includegraphics{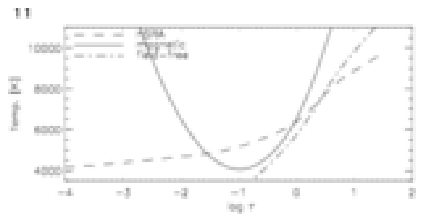}}}\fbox{\resizebox{6cm}{!}{\includegraphics{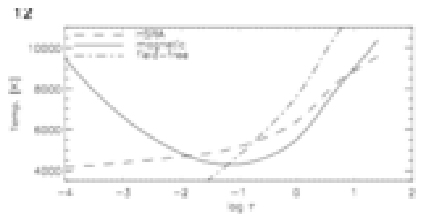}}}
\caption{Same as Fig.~\ref{specs1} for TIP Op.~005.\label{specs2}}
\end{figure*}

\begin{figure*}
\fbox{\resizebox{6cm}{!}{\includegraphics{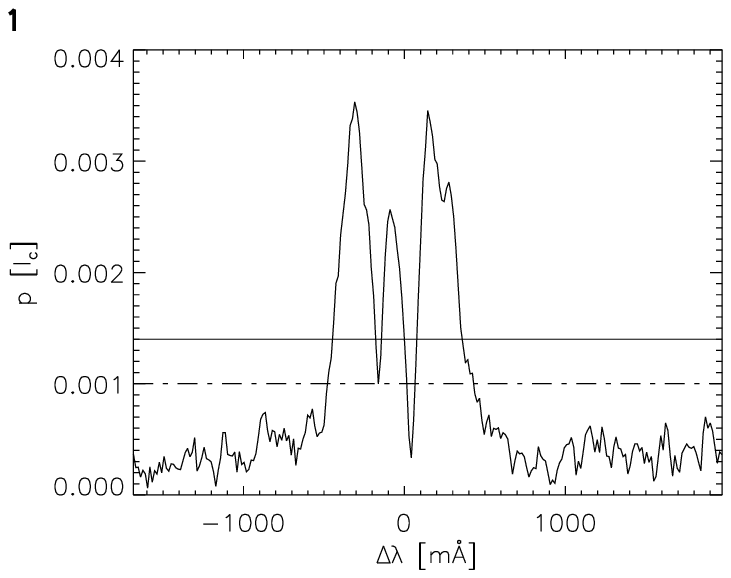}}}
\fbox{\resizebox{6cm}{!}{\includegraphics{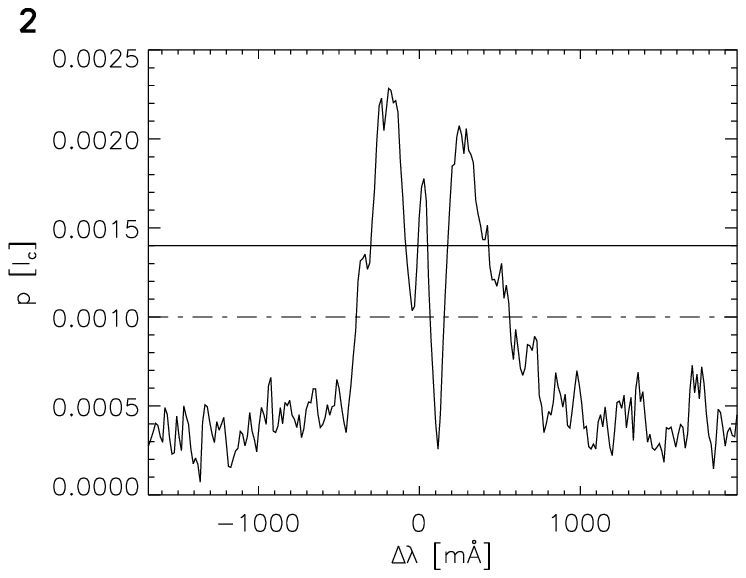}}}
\fbox{\resizebox{6cm}{!}{\includegraphics{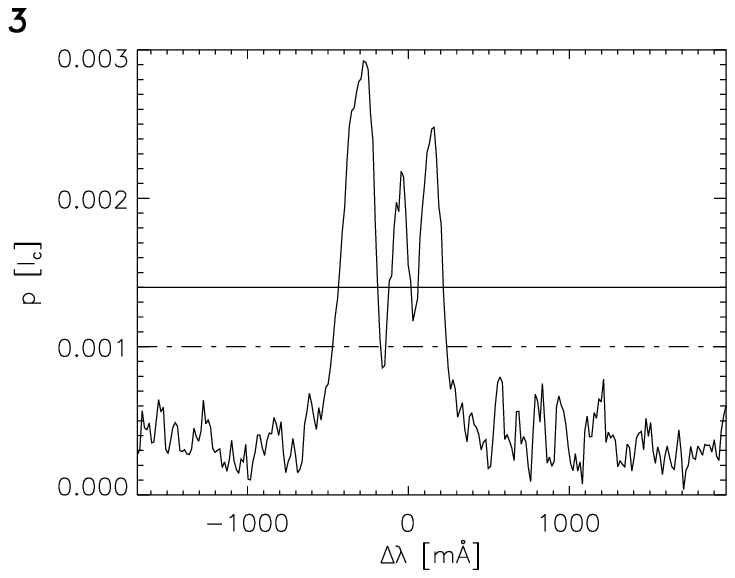}}}\\
\fbox{\resizebox{6cm}{!}{\includegraphics{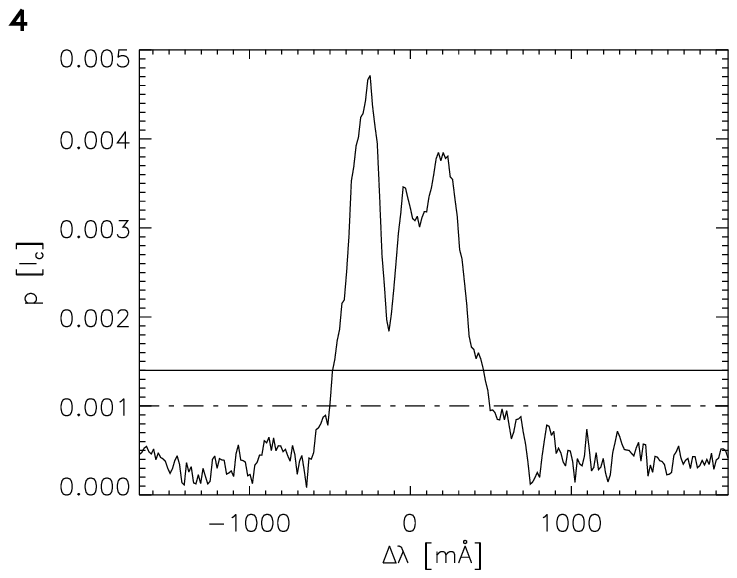}}}
\fbox{\resizebox{6cm}{!}{\includegraphics{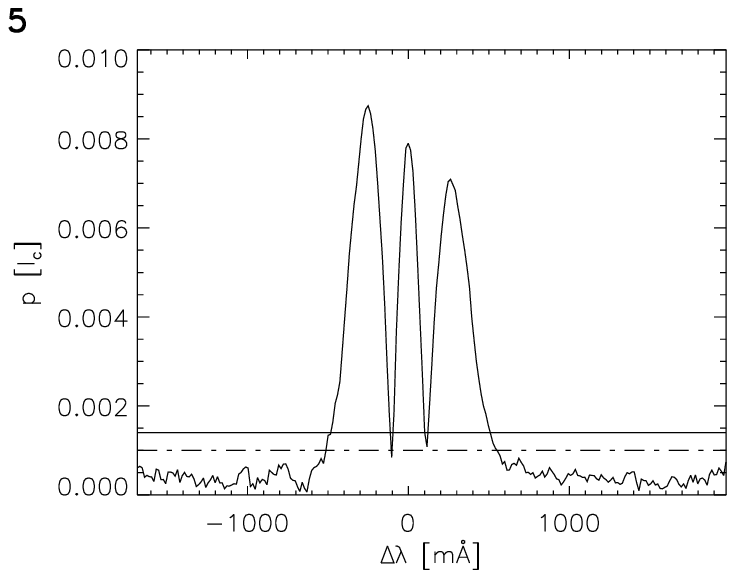}}}
\fbox{\resizebox{6cm}{!}{\includegraphics{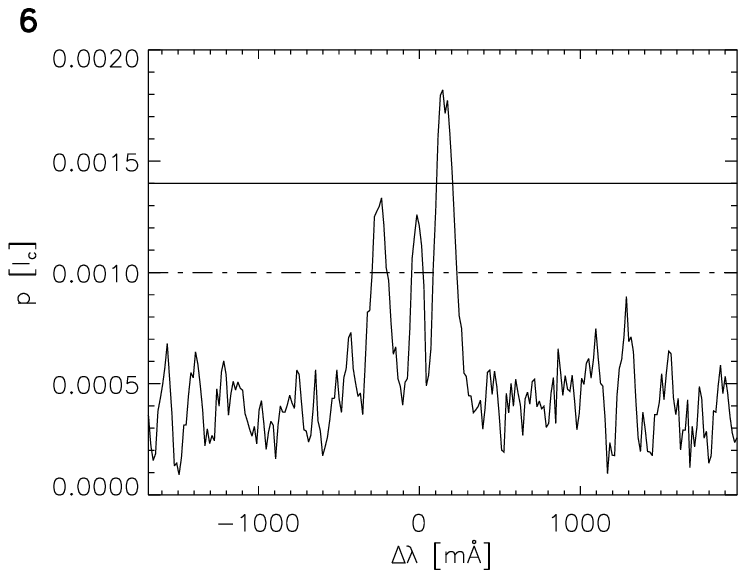}}}\\
\fbox{\resizebox{6cm}{!}{\includegraphics{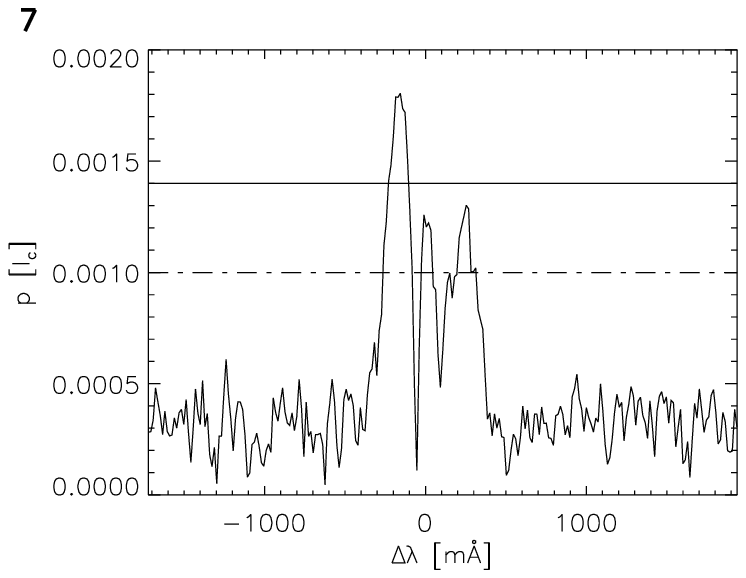}}}
\fbox{\resizebox{6cm}{!}{\includegraphics{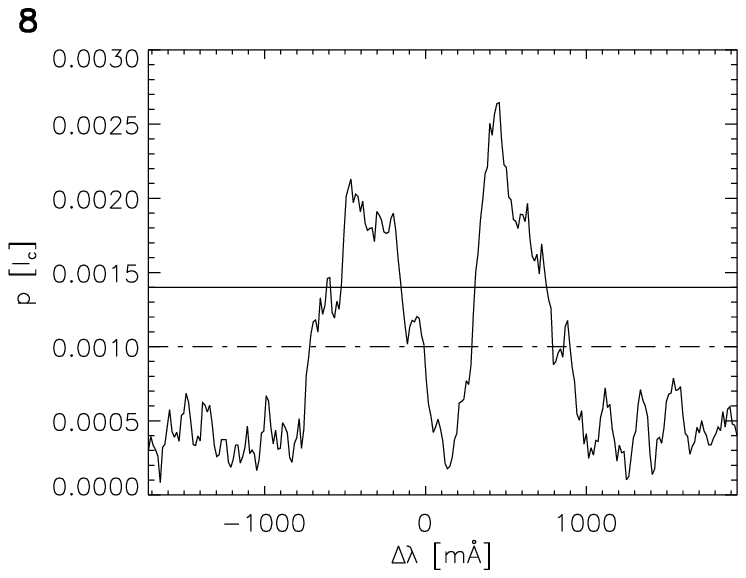}}}
\fbox{\resizebox{6cm}{!}{\includegraphics{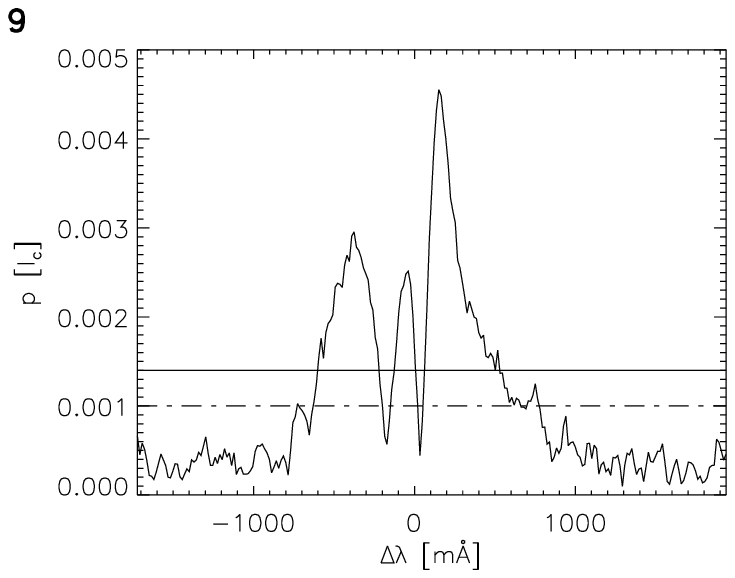}}}\\
\fbox{\resizebox{6cm}{!}{\includegraphics{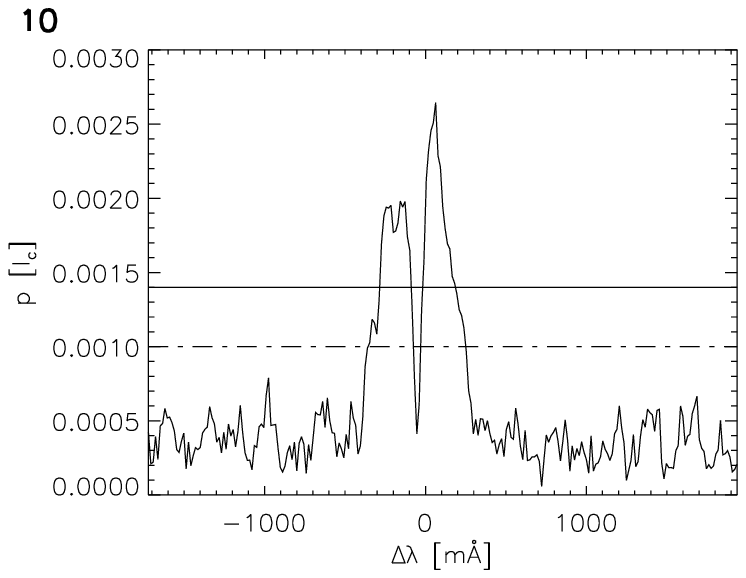}}}
\fbox{\resizebox{6cm}{!}{\includegraphics{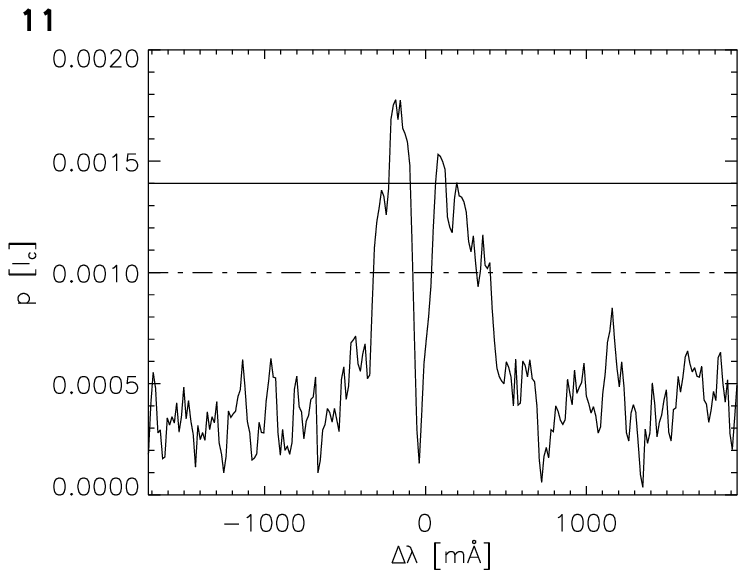}}}
\fbox{\resizebox{6cm}{!}{\includegraphics{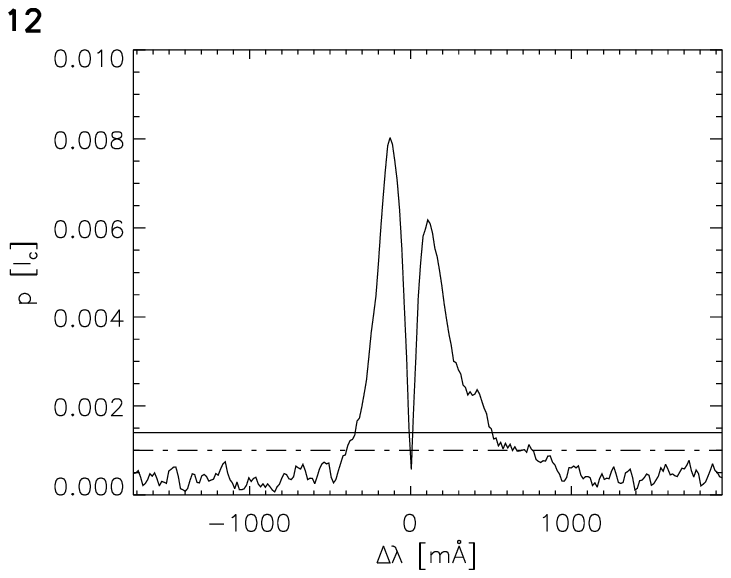}}}
\caption{Polarization degree of 1564.8 nm for the profiles shown in Fig.~\ref{specs1} ({\em upper two rows}) and \ref{specs2} ({\em lower two rows}). The {\em dash-dotted} and {\em solid} horizontal lines denote the inversion and final rejection threshold, respectively. \label{p_exams}}
\end{figure*}

Figures \ref{specs1} and \ref{specs2} show several profiles taken from the first and second long-integrated observation on 21/05/08 (Op.~001 and Op.~005), respectively. { The positions of the profiles are marked with a consecutive number in Fig.~\ref{fig1}. The profiles shown were selected to have a small polarization degree that in some cases was barely sufficient to meet the inversion threshold (e.g., profiles no.~6 and 7). Below the spectra, the temperature stratifications that were used for the generation of the best-fit spectra are shown}. With 3 nodes in temperature, the SIR code can use a parabola for changing the stratification; the parabola shape appears quite prominent for many of the locations. Note, however, that the IR lines at 1.56$\mu$m are not sensitive to the temperature in the atmosphere above log $\tau\sim -1.5$ \citep{cabrera+bellot+iniesta2005}. Only one profile corresponds to a kG field (Fig.~\ref{specs2}, {\em top middle}, { no.~8}). Figure \ref{p_exams} shows the polarization degree of 1564.8 nm for all profiles of the previous figures. { Profile no.~9} exceeds the inversion threshold of 0.001 of $I_c$ near +750 m${\AA}$ with a spike that presumably is not of solar origin, but noise in the Stokes $U$ profile. The final rejection threshold of 0.0014 is{, however}, only reached by signals clearly related to the Zeeman effect (multiple double or triple lobes).
\paragraph{Reliability of the inversion results}
{ As discussed in Sect. 3, we used a constant value for the magnetic field strength ($B$), inclination ($\gamma$), azimuth ($\psi$), and the LOS velocity. This inversion setup cannot reproduce antisymmetric Stokes $Q$ or $U$ or symmetric Stokes $V$ profiles that would require gradients in the magnetic field strength and the velocities along the LOS. The inversion was initialized with the same model atmosphere on all pixels ($B=0.9$ kG, $\psi=65$ deg), only the inclination was modified to be 10 or 170 deg depending on the polarity. In the inversion process, the equal weight used for $QUV$ in the calculation of $\chi^2$ naturally favors the component with larger polarization signal. For example, in profile no.~5 in Fig.~\ref{specs1} the $Q$ and $U$ signals are larger than the $V$ signal by almost an order of magnitude, leading to a better fit quality for $Q$ and $U$ than for $V$. In polarimetric data of lower S/N, a difference of this order usually implies that the weaker signal is not seen at all.

SIR calculates an error estimate for the free fit parameters using the diagonal elements of the covariance matrix, approximated by response functions \citep[][SIR manual]{bellotrubio+etal2000}. The error estimate depends on the number of degrees of freedom in each variable; for parameters constant with optical depth thus a single value is returned. The error estimate gives, however, only the information on how reliable the best-fit solution of the found $\chi^2$-minimum is inside the chosen inversion setup. The estimated errors for the profiles shown in Figs. 4, \ref{specs1}, and \ref{specs2} are noted on the Stokes $I$ panel of Fe\,I\,1565.2\,nm line. The average uncertainties in the calculated magnetic field strength and inclination angle given by SIR are $\pm 50\,$G and $\pm 10\,$ deg, respectively. The values agree with a previous error estimate in \citet{beck2006} that was derived from a direct analysis of the profile shape of the 1.56 $\mu$m lines (Table 3.2 on p. 47; $\delta B \sim 50$\,G and $\delta \theta \sim 5$\,deg).}
\section{Calibration of $L_{tot}$ to transversal flux \label{lin_cal}}
\begin{figure}
\includegraphics{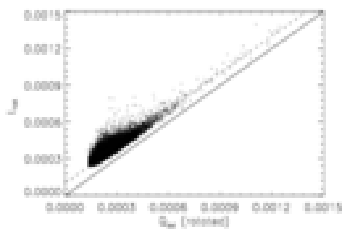}
\caption{Scatterplot of the integrated Stokes $Q$ signal in the preferred reference frame vs the total linear polarization without rotation. {\em Solid line}: unity slope; {\em dashed line}: unity slope with an offset of 0.0001.\label{lq_comp}}
\end{figure}

We tried to follow the procedure described in LI08 to obtain a calibration of linear polarization signal into transversal magnetic flux that is independent of the inversion results. To reduce the influence of noise, the latter authors first determine the ``preferred azimuth frame'' where the linear polarization signal is concentrated in Stokes $Q$. To achieve this, we determined the azimuth angle from the ratio of $U$ and $Q$ and rotated the spectra correspondingly to maximize the Stokes $Q$ signal. The scatterplot in Fig.~\ref{lq_comp} compares the previously used total linear polarization, $L_{tot} = \int \sqrt{Q^2+U^2}(\lambda) d\lambda$, with the corresponding $Q_{tot}({\rm rotated}) = \int |Q(\lambda)|d\lambda$ as a measure of the linear polarization. The rotation of the spectra reduces the noise contribution by a constant amount, but the old and new values otherwise have a linear relationship with a slope close to unity.

\begin{figure}
\resizebox{7.5cm}{!}{\includegraphics{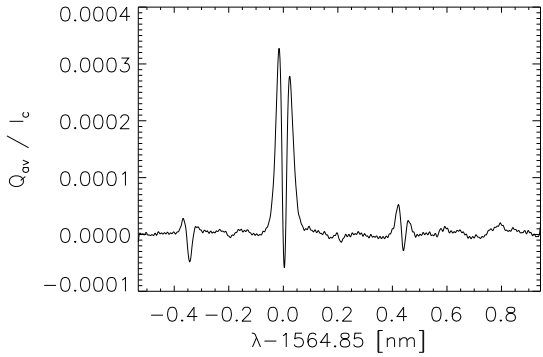}}
\caption{The average Stokes $Q$ profile.\label{qmask}}
\end{figure}

We then averaged the rotated $Q$ spectra over all spatial positions exceeding the polarization threshold for the inversion. The average Stokes $Q$ spectrum is used as a spectral mask by LI08, but unfortunately their method fails for the infrared lines. The wavelengths around the line core have negative values in the average $Q$ profile (Fig.~\ref{qmask}) which prevents using it in the same way as in LI08. We thus used $Q_{tot}({\rm rotated})$ as defined above instead which we think to be equivalent to that of LI08 despite of not using a (somewhat arbitrary) spectral mask. 

\begin{figure}
\includegraphics{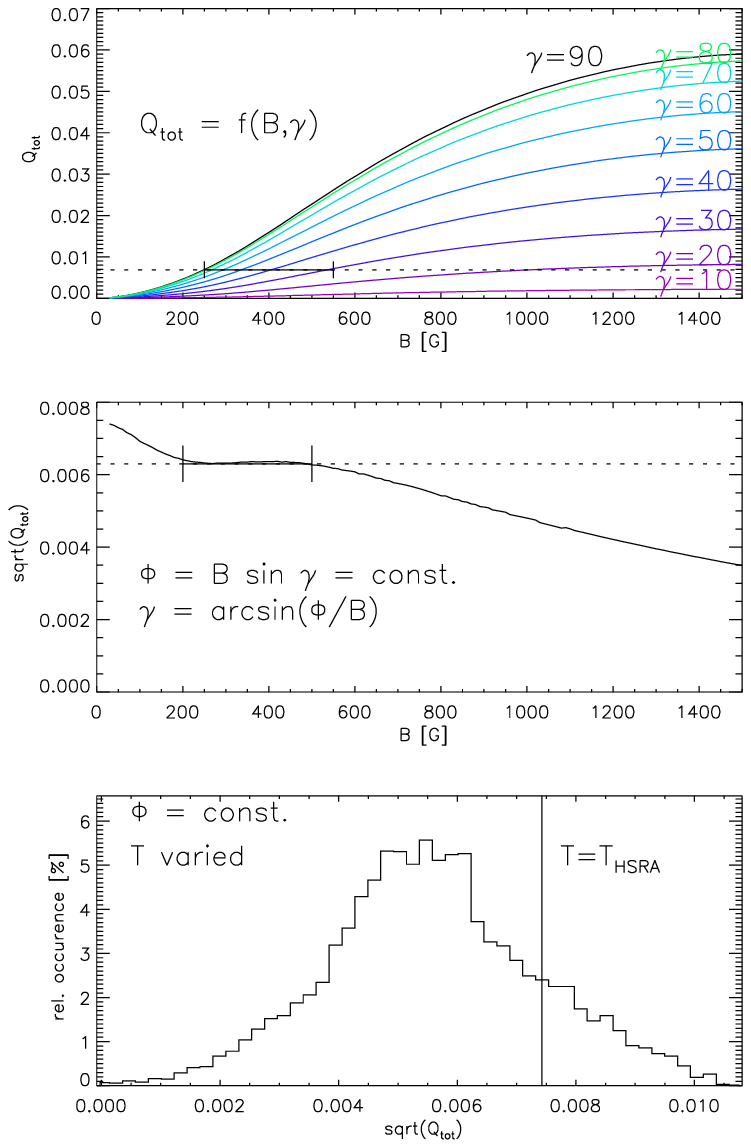}
\caption{{\em Top}: calibration curves from $Q_{tot}$ into field strength $B$ for field inclinations $\gamma$ from 10 to 90 deg ({\em bottom to top}). The {\em horizontal dotted line} is at $Q_{tot}=0.007$; the {\em solid part} of it denotes a range in $B$ that gives the same $Q_{tot}$ at different $\gamma$. {\em Middle}: $\sqrt{Q_{tot}}$ vs field strength for constant magnetic flux. {\em Dotted line} and {\em solid part} as above for $\sqrt{Q_{tot}}=0.0063$. {\em Bottom}: histogram of $\sqrt{Q_{tot}}$ for constant flux but varying temperature stratifications $T$. The {\em vertical line} denotes the value resulting from the HSRA atmosphere model. \label{qcal}}
\end{figure}

The plot of $Q_{tot}({\rm rotated})$ vs transversal flux (Fig.~\ref{fig7}, {\em middle upper panel}), showed considerable scatter that already put the use of a single calibration curve into doubt. We thus not only tried to obtain a calibration curve, but also to quantify the effect of various parameters on the obtained relation. The {\em upper part} of Fig.~\ref{qcal} shows calibration curves of $Q_{tot}$ vs field strength for different field inclinations $\gamma$. The uppermost curve for $\gamma = 90^\circ$ corresponds to the one used by LI08. With the assumption that the field inclination not necessary equals $90^\circ$, one already finds that one and the same value of $Q_{tot}$ can be obtained in a range of around 200-550 G in field strength. The same effect is shown in the {\em middle part} where the magnetic flux, $\Phi=B\sin \gamma$, was kept constant at 1.8$\cdot 10^{16}$ Mx, $B$ was varied, and $\gamma$ was derived accordingly from $\gamma={\rm arcsin} (\Phi/B)$. Again a range of around 200-500 G in $B$ gives the same value of $Q_{tot }$. As final test we chose to investigate the influence of the temperature stratification on the resulting $Q_{tot}$-value. We kept magnetic flux, field strength and field inclination constant at (1.8$\cdot 10^{16}$ Mx, 20 G, $75^\circ$), and synthesized spectra for different temperature stratifications. We used 10000 temperature stratifications that were derived for the magnetic component in the inversion, and thus can be taken to be an estimate for the range of temperatures to be expected in the quiet Sun. The histogram of the resulting $Q_{tot}$-values is displayed in the {\em bottom part} of Fig.~\ref{qcal}. $\sqrt{Q_{tot}}$ ranges from nearly zero up to 0.01, which roughly also corresponds to the scatter of $\sqrt{Q_{tot}}$ in Fig.~\ref{fig7}. We thus think that the biggest contribution for the scatter comes from temperature effects. We remark that we used a magnetic filling factor of unity in all calculations. Any additional variation of the filling factor for unresolved magnetic structures will increase the scatter of $Q_{tot}$ even more.

We conclude that the usage of a calibration curve for a derivation of transversal magnetic flux from $L_{tot}$ or $Q_{tot}$, regardless of the exact calculation of the wavelength integrated quantities, is not reliable for a solid estimate, mainly because the strong influence of the thermodynamical state of the atmosphere on the weak polarization signals.
\clearpage
\clearpage
\end{appendix}
\end{document}